\documentclass[lettersize,journal]{IEEEtran}
\usepackage{amsmath,amsfonts}
\usepackage{algorithmic}
\usepackage{algorithm}
\usepackage{array}
\usepackage{setspace}
\usepackage[caption=false,font=normalsize,labelfont=sf,textfont=sf]{subfig}
\usepackage{textcomp}
\usepackage{stfloats}
\usepackage{url}
\usepackage{changes}
\usepackage{verbatim}
\usepackage{graphicx}
\usepackage{cite}
\usepackage{stfloats}
\usepackage{textcomp}
\usepackage{stfloats}
\usepackage{url}
\usepackage{verbatim}
\usepackage{graphicx}
\usepackage{cite}
\usepackage{booktabs}
\usepackage{multirow}
\begin{document}
	
	\title{Graph Embedding with Mel-spectrograms for Underwater Acoustic Target Recognition}
	
	\author{Sheng Feng, Shuqing Ma, Xiaoqian Zhu
		\thanks{This paper was produced by the IEEE Publication Technology Group. (corresponding author: Xiaoqian Zhu.)}
		\thanks{Manuscript received May 5, 2024; This work was supported by the National Defense Fundamental Scientific Research Program under Grant No.JCKY2020550C011. The authors are with the College of Meteorology and Oceanography, National University of Defense Technology, Chang sha 410073, China (e-mail: fengsheng18@nudt.edu.cn; mashuqing@nudt.edu.cn;
			zhu\_xiaoqian@sina.com).}}
	
	\markboth{Journal of \LaTeX\ Class Files,~Vol.~14, No.~8, August~2021}%
	{Shell \MakeLowercase{\textit{et al.}}: A Sample Article Using IEEEtran.cls for IEEE Journals}
	
	\IEEEpubid{0000--0000/00\$00.00~\copyright~2021 IEEE}
	
	\maketitle
	
	\begin{abstract}
		Underwater acoustic target recognition (UATR) is extremely challenging due to the complexity of ship-radiated noise and the variability of ocean environments. Although deep learning (DL) approaches have achieved promising results, most existing models implicitly assume that underwater acoustic data lie in a Euclidean space. This assumption, however, is unsuitable for the inherently complex topology of underwater acoustic signals, which exhibit non-stationary, non-Gaussian, and nonlinear characteristics. To overcome this limitation, this paper proposes the UATR-GTransformer, a non-Euclidean DL model that integrates Transformer architectures with graph neural networks (GNNs). The model comprises three key components: a Mel patchify block, a GTransformer block, and a classification head. The Mel patchify block partitions the Mel-spectrogram into overlapping patches, while the GTransformer block employs a Transformer Encoder to capture mutual information between split patches to generate Mel-graph embeddings. Subsequently, a GNN enhances these embeddings by modeling local neighborhood relationships, and a feed-forward network (FFN) further performs feature transformation. Experiments results based on two widely used benchmark datasets demonstrate that the UATR-GTransformer achieves performance competitive with state-of-the-art methods. In addition, interpretability analysis reveals that the proposed model effectively extracts rich frequency-domain information, highlighting its potential for applications in ocean engineering.
	\end{abstract}

	
	\section{Introduction}
	\IEEEPARstart{U}{nderwater} acoustic target recognition (UATR), a crucial topic in ocean engineering, involves detecting and classifying underwater targets based on their unique acoustic properties. This capability holds important implications for maritime security, environmental monitoring, and underwater exploration. However, UATR is highly challenging due to the complex mechanisms of underwater sound propagation in diverse marine environments \cite{xie2022adaptive}. Factors such as attenuation, scattering, and reverberation significantly complicate target identification and classification. Early UATR methods primarily relied on experienced sonar operators for manual recognition, but such approaches are prone to subjective influences, including psychological and physiological conditions. To overcome these limitations, statistical learning techniques were introduced, leveraging time-frequency representations derived from waveforms to enhance automatic recognition. Representative approaches include Support Vector Machines (SVM) \cite{7435957,7108260} and logistic regression \cite{10390008}. Nevertheless, as the demand for higher recognition accuracy has increased, the shortcomings of statistical learning-based methods have become apparent. These methods typically capture only shallow discriminative patterns and fail to fully exploit the potential of diverse datasets.
	
	Deep learning (DL), as a subset of machine learning, has achieved remarkable progress in UATR by learning complex patterns from large volumes of acoustic data \cite{yang2020underwater,10.1121/1.5133944}. Among DL models, convolutional neural networks (CNNs) have been widely studied for end-to-end modeling of acoustic structures, owing to their strong feature extraction capabilities. For example, \cite{doan2020underwater} proposed a dense CNN that outperformed traditional methods by extracting meaningful features from waveforms. Similarly, \cite{sun2022underwater} employed ResNet and DenseNet to identify synthetic multitarget signals, demonstrating effective recognition of ship signals using acoustic spectrograms. A separable and time-dilated convolution-based model for passive UATR was proposed in \cite{s21041429}, showing notable improvements over conventional approaches. In addition, \cite{liu2021underwater} introduced a fusion network combining CNNs and recurrent neural networks (RNNs), achieving strong recognition performance across multiple tasks through data augmentation. Despite these successes, the inherent local connectivity and parameter-sharing properties of CNNs bias them toward local feature extraction, making it difficult to capture global structures such as overall spectral evolution and relationships among key frequency components. \IEEEpubidadjcol
	
	To address this issue, attention mechanisms have been integrated into DL models to capture long-range dependencies in acoustic signals \cite{10012335}. For instance, \cite{xiao2021underwater} proposed an interpretable neural network incorporating an attention module, while \cite{ZHOU2023115784} designed an attention-based multi-scale convolution network that extracted filtered multi-view representations from acoustic inputs and demonstrated effectiveness on real-ocean data. Leveraging the Transformer’s multi-head self-attention (MHSA) mechanism, \cite{feng2022transformer} proposed a lightweight UATR-Transformer, which achieved competitive results compared to CNNs. Inspired by the Audio Spectrogram Transformer (AST) \cite{DBLP:conf/interspeech/GongCG21}, a spectrogram-based Transformer model (STM) was applied to UATR \cite{jmse10101428}, yielding satisfactory outcomes. Moreover, self-supervised Transformers have shown strong potential in extracting intrinsic characteristics of underwater acoustic data \cite{10.1121/10.0015053,10414073,10.1121/10.0019937}. Nonetheless, the complexity of pre-training and the unclear internal mechanisms suggest that this line of research is still in its early stages. In summary, current UATR research primarily focuses on extracting discriminative features through convolution, attention, and their variants \cite{tian2023joint,YANG2024107983}, which have achieved encouraging results with promising applications.
	
	In practice, underwater acoustic data are often regarded as high-dimensional topological data due to their irregular structure and cluttered characteristics \cite{esfahanian2013using}. The generation and radiation of underwater target noise involve multiple components, including broadband continuous spectra, strong narrowband lines, and distinct modulation features. As a result, underwater signals often exhibit nonlinear, non-stationary, and non-Gaussian behavior. In the time domain, the waveforms and amplitudes vary dynamically, while in the frequency domain, spectral distributions can change over time. These characteristics challenge the representation of acoustic features as simple Euclidean vectors. Traditional models directly process sequential Euclidean data, such as images or audio, focusing on optimizing local and global information extraction. However, they neglect the geometric structure of acoustic data in high-dimensional space and overlook the non-Euclidean nature of the signals, leading to suboptimal performance.
	
	To address this limitation, we propose the UATR-GTransformer, a non-Euclidean DL model that performs recognition via Mel-graph embeddings. The motivation for graph modeling on the Mel-spectrogram stems from the strength of graph theory in handling complex structures and uncovering latent patterns in topological data \cite{Waikhom2023}, thereby providing a promising solution to the challenges of non-stationarity, non-Gaussianity, and nonlinearity \cite{7763882,9526764,PhysRevE.92.022817}. In the proposed framework, the acoustic signal is first transformed into a Mel-spectrogram and partitioned into overlapping patches. A Transformer Encoder then extracts features, capturing global dependencies via MHSA to form Mel-graph embeddings. Each embedding is subsequently treated as a graph node, and edges are defined by relationships among nodes. This Mel-graph captures both local and global structures of the spectrogram, enabling the discovery of hidden patterns. Through further graph processing, it is expected that the UATR-GTransformer can effectively exploit the topological structure of acoustic features to enhance recognition performance.
	
	The main contributions of this paper are as follows:
	\begin{itemize}
		\item We propose a non-Euclidean framework for intelligent UATR that explicitly incorporates spatial information from acoustic features. To the best of our knowledge, this is the first work to introduce graph structures into UATR. Mel-graph processing enables the model to leverage topological characteristics of underwater acoustic signals.
		\item We integrate a Transformer Encoder to enhance global feature perception during graph processing. By propagating global information across neighboring nodes, the graph representation becomes more robust.
		\item We provide interpretability through attention and graph visualization, allowing better understanding of the prediction process and increasing the model’s practicality for ocean engineering applications.
	\end{itemize}
	
	
	\section{Gaussianity and Linearity Test}
	In this section, we examine the Gaussianity and linearity of sonar-received radiated noise using Hinich theory \cite{Hinich1982}, which provides an effective framework to validate the non-Gaussian and nonlinear characteristics of random processes. 
	
	Let $x$ denote the ship-radiated noise with probability density function $f(x)$. Its moment generating function (MGF) can be defined as:
	\begin{equation}
		\Phi(\omega)=\int_{-\infty}^{\infty} f(x) e^{j \omega x} \mathrm{~d} x.
	\end{equation}
	The $k$-th order moment is obtained by differentiating $\Phi(\omega)$ $k$ times with respect to $\omega$:
	\begin{equation}
		m_k=\left.(-j)^k \frac{\mathrm{d}^k \Phi(\omega)}{\mathrm{d} \omega^k}\right|_{\omega=0}.
	\end{equation}
	Based on the relationship between the cumulant generating function and the MGF, $\Psi(\omega)=\ln \Phi(\omega)$, the $k$-th order cumulant is expressed as:
	\begin{equation}
		c_k=\left.(-j)^k \frac{\mathrm{d}^k \Psi(\omega)}{\mathrm{d} \omega^k}\right|_{\omega=0}.
	\end{equation}
	According to Hinich theory, if the third-order cumulants of a process are zero, its bispectrum and bicoherence are also zero, indicating Gaussianity. Conversely, a nonzero bispectrum implies that the process is non-Gaussian.
	
	The hypothesis testing can be formulated as follows: the null hypothesis $\mathbf{H_0}$ assumes that the underwater acoustic signal is Gaussian, i.e., its higher-order cumulants are zero; the alternative hypothesis $\mathbf{H_1}$ assumes the opposite, i.e., the signal is non-Gaussian. The probability of false alarm (PFA) reflects the risk of incorrectly accepting $\mathbf{H_1}$. Typically, if $\mathrm{PFA} \geq 0.05$, $\mathbf{H_0}$ is accepted; whereas when $\mathrm{PFA} \to 0$, $\mathbf{H_1}$ is accepted. To further assess nonlinearity, a comparison between the theoretical and estimated interquartile deviations is conducted. A large deviation suggests nonlinearity, while a small deviation indicates linearity. 
	\begin{figure}[h]
		\centering
		\begin{minipage}{1\linewidth}
			\centerline{\includegraphics[width=0.98\textwidth]{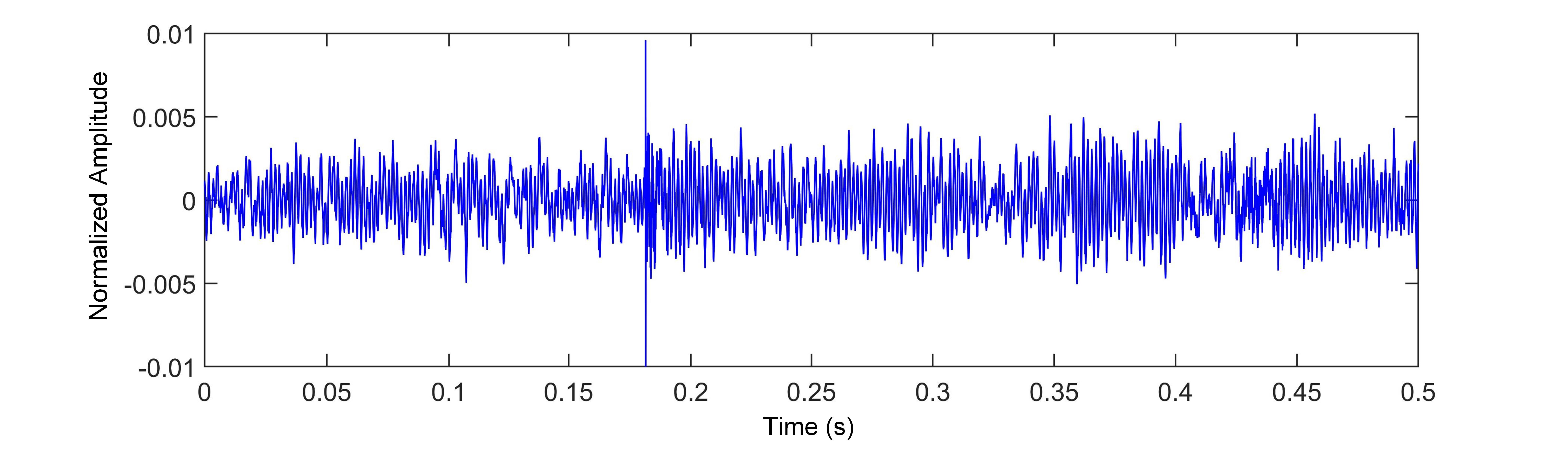}}
			\centerline{(a)}
		\end{minipage}
		\begin{minipage}{1\linewidth}
			\centerline{\includegraphics[width=0.98\textwidth]{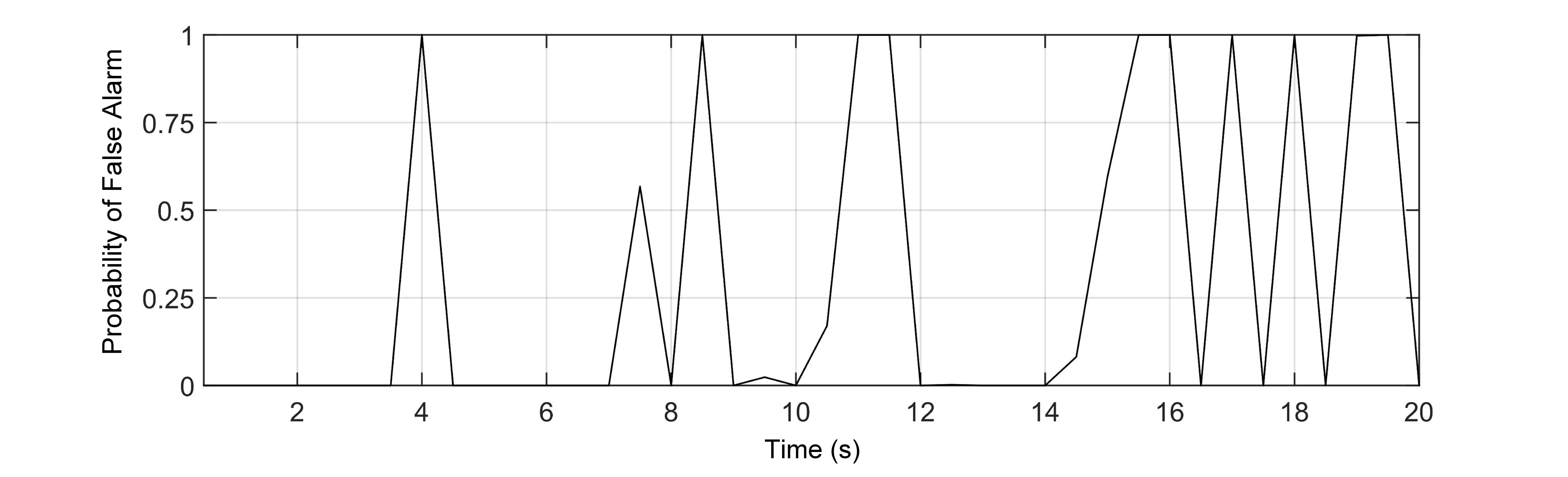}	}
			\centerline{(b)}	
		\end{minipage}
		\begin{minipage}{1\linewidth}
			\centerline{\includegraphics[width=0.98\textwidth]{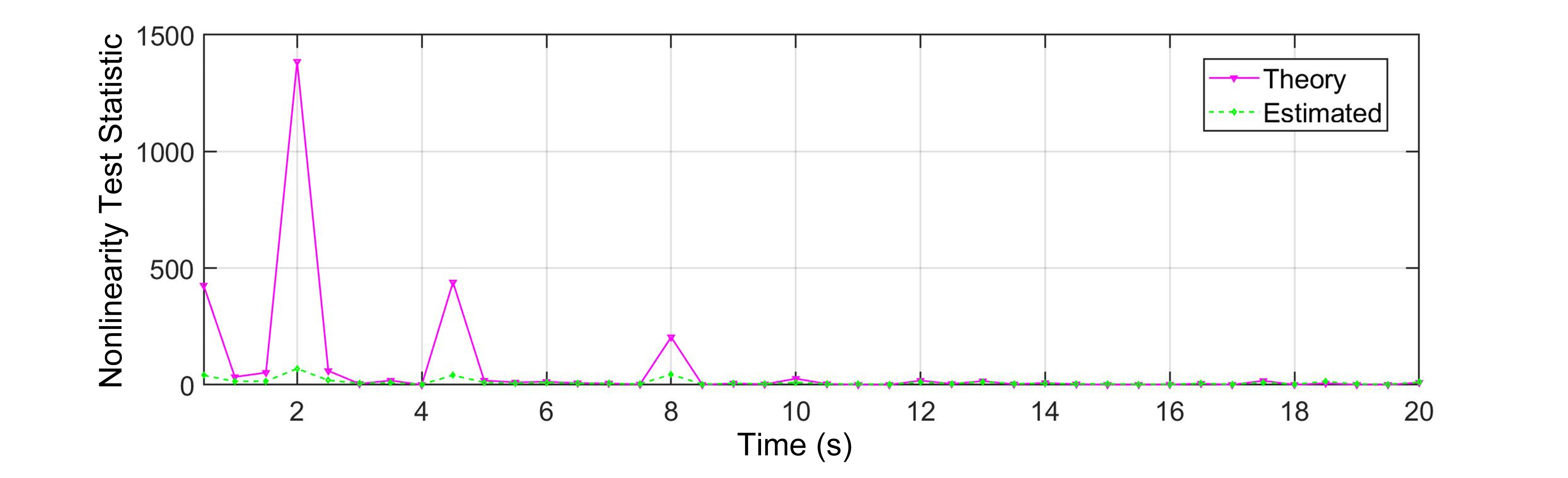}}
			\centerline{(c)}
		\end{minipage}
		\caption{Hinich hypothesis testing on the ShipsEar dataset: (a) waveform of one segment; (b) Gaussianity test results; (c) linearity test  results. 
		}
		\label{fig:hinich}
	\end{figure}
	\begin{figure*}[ht]
		\begin{minipage}{0.5\linewidth}
			\centerline{\includegraphics[width=0.688\textwidth]{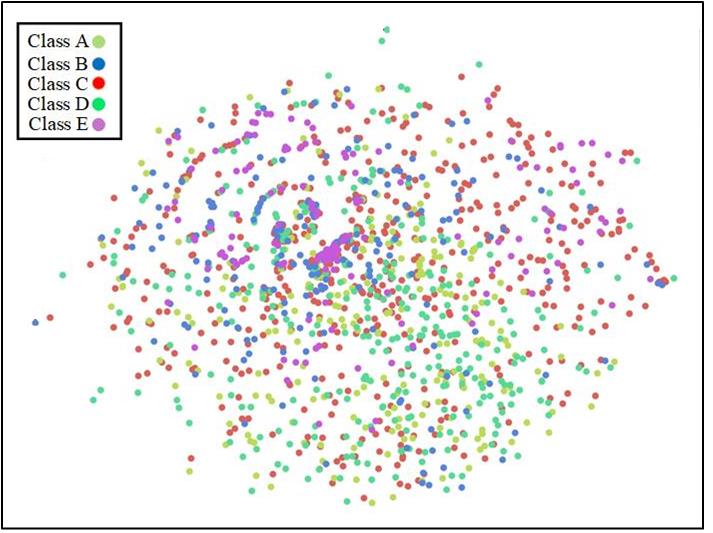}}
			\centerline{(a)}
		\end{minipage}
		\begin{minipage}{0.5\linewidth}
			\centerline{\includegraphics[width=0.688\textwidth]{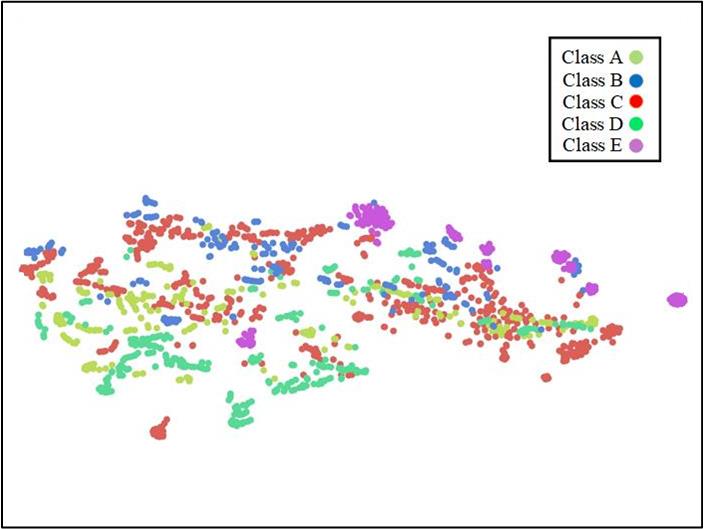}}
			\centerline{(b)}
		\end{minipage}
		\caption{Topological structure of the ShipsEar dataset using the t-SNE algorithm \cite{JMLR:v9:vandermaaten08a}. (a) waveform distribution; (b) Mel-Fbank feature distribution.}
		\label{fig:topological}
	\end{figure*}
	
	Fig.~\ref{fig:hinich} presents the Hinich test results based on a 20-s sample selected from the ShipsEar dataset \cite{santos2016Shipsear}, implemented using the HOSA package \cite{Swami2025}. The original sampling frequency of the signal is 52374 Hz, and it was segmented into 40 intervals of 0.5 s each for Gaussianity and linearity evaluation. Previous studies have already demonstrated the non-stationary characteristic of underwater acoustic signals \cite{10.1121/10.0003382,10.1121/1.4776775}. As shown in Fig.~\ref{fig:hinich}(b), the PFA values of the Gaussianity test vary between 0 and 1. In particular, multiple instances exhibit $\mathrm{PFA}=0$, indicating strong non-Gaussianity. Moreover, the significant deviation between the estimated and theoretical interquartile ranges further confirms nonlinearity. 
	Following t-SNE visualization using the HyperTools package \cite{hypertools} with default parameters, Fig.~\ref{fig:topological} clearly illustrates that both the waveform and the time-frequency representation of underwater acoustic signals exhibit complex structures, forming high-dimensional topological patterns in a non-Euclidean space. Notably, the time-frequency features demonstrate better class separability than raw waveforms, validating their effectiveness for underwater target classification.

	\section{Proposed Method}
\begin{figure*}
		\centering
		\includegraphics[width=15.0cm]{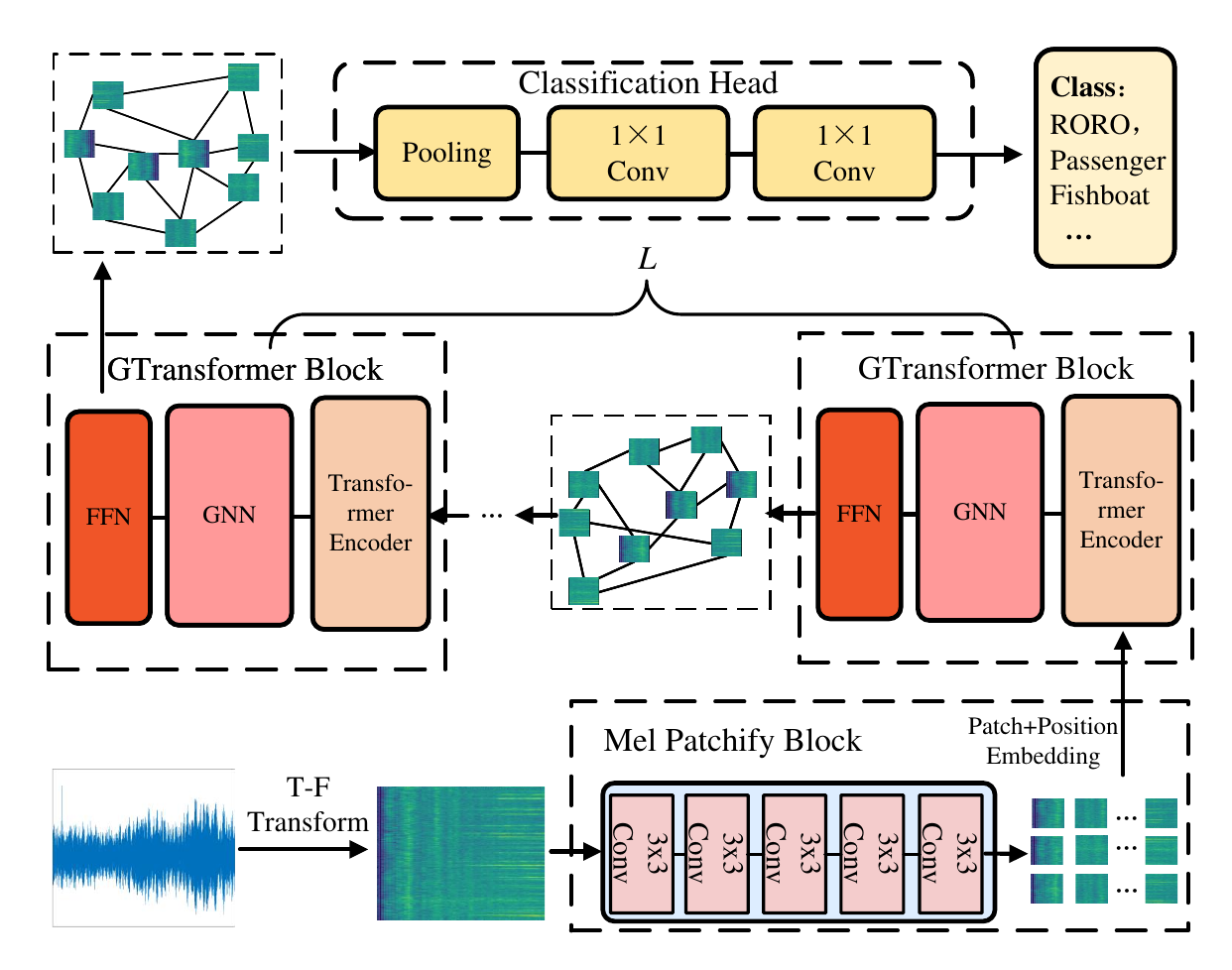}
		\caption{Overall workflow of the proposed UATR-GTransformer framework.}
		\label{fig:uatrgt}
	\end{figure*}

	For UATR in topological space, we propose a Mel-graph embedding-based DL model to recognize real-world underwater acoustic signals. The overall framework is illustrated in Fig.~\ref{fig:uatrgt}, which comprises four main components: Mel-spectrogram feature extraction, the Mel Patchify Block, the GTransformer Block, and a classification head. In this section, we first describe the extraction of Mel-spectrogram features, followed by the partitioning of the spectrogram using the Mel Patchify Block. The construction and updating of the Mel-graph are performed within the GTransformer Block. Finally, we provide a brief overview of the classification head.
	
	\subsection{Mel-spectrogram Feature}
	In the context of UATR, the Mel-spectrogram, derived from the Mel filterbank (Mel-Fbank), has become a widely adopted time–frequency representation in sonar signal processing \cite{liu2021underwater}. In this work, the choice of Mel-spectrograms as model input is motivated by their partially overlapping frequency bands, which preserve intrinsic signal information and exhibit high inter-feature correlation. Consequently, when further processed through graph modeling, the connections among graph nodes are strengthened, enabling the construction of a more discriminative topological graph.
	
	The extraction of Mel-spectrogram features involves the following steps, after resampling the input signal to 16 kHz:
	
	(1) \textbf{Pre-emphasis:} This step enhances the energy of high-frequency components for spectrum balancing. It is typically implemented by processing the original signal $x[n]$ as follows:
	\begin{equation}
		y[n] = x[n] - \alpha x[n-1],
	\end{equation}
	where $y[n]$ is the pre-emphasized signal and $\alpha$ is the pre-emphasis coefficient, usually set to $0.97$, approximated by a hardware-friendly coefficient \cite{10.1007/978-981-99-7505-1_61}.
	
	(2) \textbf{Framing:} The pre-emphasized signal $y[n]$ is segmented into overlapping frames, each containing 25 ms of audio with a frame shift of 10 ms.
	
	(3) \textbf{Windowing:} To mitigate spectral leakage, each frame is multiplied by a Hanning window.
	
	(4) \textbf{Fast Fourier Transform (FFT):} The FFT is then applied to each windowed frame to transform the signal into its frequency-domain representation.
	
	(5) \textbf{Mel Filtering:} The frequency-domain signal is filtered using a 128-band triangular Mel-Fbank, defined as
	\begin{equation}
		F_m(k)= \begin{cases}0 & \text { if } k<f[m-1], \\[4pt]
			\frac{k-f[m-1]}{f[m]-f[m-1]} & \text { if } f[m-1] \leq k<f[m], \\[6pt]
			\frac{f[m+1]-k}{f[m+1]-f[m]} & \text { if } f[m] \leq k<f[m+1], \\[6pt]
			0 & \text { if } k \geq f[m+1], \end{cases}
	\end{equation}
	where $f[i]$ denotes the $i$-th center frequency of the Mel bins and $k$ is the frequency index. The filterbank energy is then applied to the Short-Time Fourier Transform (STFT) coefficient $X(k)$ to compute the Mel-spectrogram:
	\begin{equation} 
		M = \log\left(\sum_{k=0}^{N-1 } F_m(k) \times X(k)\right),
	\end{equation}
	where $N=128$ is the number of Mel frequency bins. The above extraction procedure is implemented using the \textit{torchaudio} package. Suppose the received underwater acoustic signal has a duration of 5 s, the resulting Mel-spectrogram will have a dimension of $512 \times 128$ after time padding.
	
	\subsection{Mel Patchify Block}
	Previous studies have shown that patch modeling of acoustic spectrograms can effectively capture meaningful time–frequency structures from acoustic signals \cite{gong2022ssast}. Therefore, the Mel-spectrogram is first divided into overlapping patches, which serve as the basic computational units of the model. This enables the UATR-GTransformer to construct a graph that preserves spatial information in both the time and frequency domains. Specifically, an input Mel-spectrogram is partitioned into $N$ patches of size $16 \times 16$ using the Mel patchify block. This block employs a stem convolution consisting of a sequence of trainable $3 \times 3$ convolutional kernels sliding across the spectrogram. Such convolutions are effective for extracting fine-grained features and have been shown to maintain optimization stability and computational efficiency \cite{10.5555/3540261.3542586}. In our implementation, five convolutional kernels are used to process the Mel-spectrogram. The primary objective is to extract salient features from the split patches and provide rich representations for subsequent network layers.
	
	Among these convolutional kernels, the first four use a stride of 2, while the final kernel uses a stride of 1. The stride configuration serves two purposes. The initial strides of 2 progressively downsample the feature maps to capture coarse-grained features and reduce computational cost, whereas the final stride of 1 maintains the spatial resolution for detailed representation. To further improve training stability and introduce nonlinearity, batch normalization and ReLU activation are applied after each convolutional operation. Assuming the input Mel-spectrogram size is $512 \times 128$, the resulting patch embedding has a dimension of $(dim, 32, 8)$ due to the strides of 2, 2, 2, 2, and 1. Here, $dim$ denotes the output channel size of the last convolutional kernel, which is also the graph embedding dimension.
	
	Since graph-structured representations rely on precise spatial information, a two-dimensional positional embedding is added to the patch embeddings, similar to the Transformer framework \cite{gong21b_interspeech}. This embedding captures the order of time–frequency distributions, thereby enhancing the model’s ability to process graph structures:
	\begin{equation}\label{eq:abs-pos}
		\centering
		\mathbf{x}_i \leftarrow \mathbf{x}_i + PE_i,
	\end{equation}
	where $\mathbf{x}_i$ denotes the patch embedding. Specifically, a learnable positional encoding $PE_i \in \mathbb{R}^{32 \times 8}$ is added along both the frequency and time axes of the split patches, followed by a broadcasting operation. Finally, the set of patch embeddings $\mathbf{X_0}$ is reshaped into $(256, dim)$ as input to the GTransformer Block.

	\subsection{GTransformer Block}
	As the backbone of the UATR-GTransformer, the GTransformer block consists of a Transformer Encoder, a graph neural network (GNN), and a feed-forward network (FFN).
	
	\subsubsection{Transformer Encoder}
\begin{figure}
		\centering
		\includegraphics[width=5.cm]{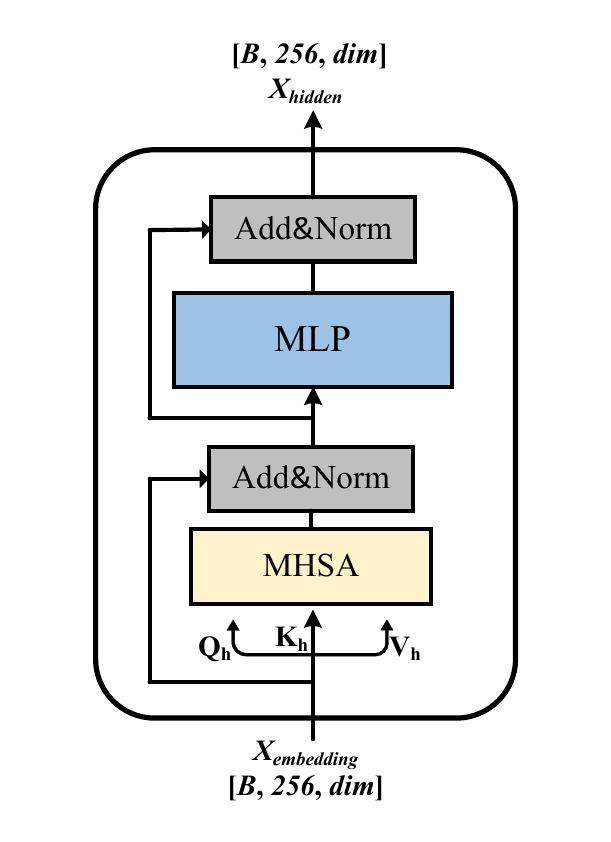}
		\caption{Illustration of the Transformer Encoder for global feature extraction. Here, $B$ denotes the batch size.}
		\label{fig:transformer}
		\vspace{-1.5em}
	\end{figure}

	In the UATR-GTransformer, the Transformer Encoder functions as a global feature extractor on $\mathbf{X}$, capturing the overall time–frequency structure. Its architecture is illustrated in Fig.~\ref{fig:transformer}. The core mechanism of the Transformer Encoder is MHSA, which projects the input features into multiple sets of queries, keys, and values. Attention is then computed independently in each head, enabling the model to capture high-level dependencies from multiple perspectives.
	The MHSA formulation for embeddings at the $l$-th layer $\mathbf{X}_{l}$ is given by:
	\begin{equation}
		\label{eq:mhsa}
		\begin{gathered}
			\mathbf{Q}_{h}, \mathbf{K}_{h}, \mathbf{V}_{h}=\mathbf{X}_{l}\mathbf{W}_{h}^{Q}, \mathbf{X}_{l}\mathbf{W}_{h}^{K}, \mathbf{X}_{l}\mathbf{W}_{h}^{V}, \\
			\operatorname{Attn}\left(\mathbf{Q}_{h}, \mathbf{K}_{h}, \mathbf{V}_{h}\right)=\operatorname{softmax}\left(\frac{\mathbf{Q}_{h} \mathbf{K}_{h}^{T}}{\sqrt{D_{\text{attn}}}}\right)\mathbf{V}_{h},
		\end{gathered}
	\end{equation}
	where $\mathbf{W}_{h}^{Q}$, $\mathbf{W}_{h}^{K}$, and $\mathbf{W}_{h}^{V}$ are learnable projection matrices for the query, key, and value sets, respectively. $H$ denotes the number of heads, $h \in [1,H]$ indexes the head, and $D_{\text{attn}}=dim/H$ is the dimensionality per head.
	
	The outputs of all $H$ attention heads, each of size $(256, dim/H)$, are concatenated to generate an attention representation of size $(256, dim)$. This representation is then passed through a multi-layer perceptron (MLP) comprising two linear layers with a GELU activation in the middle. Residual connections are applied after both the MHSA and MLP modules. Following standard Transformers, layer normalization is employed between layers instead of batch normalization to improve gradient stability and convergence.

	\subsubsection{GNN}
	In topological data processing, graphs naturally represent associative relationships among entities \cite{10530642,TORRES2024111268}. GNNs are well suited to capture and exploit these relationships by integrating node-specific features with the graph structure. Through message passing along edges, GNNs effectively learn dependencies between nodes, enabling the processing of high-dimensional topological data. In the proposed framework, a GNN is employed to construct and update the Mel-graph following the Transformer Encoder. Coupling a GNN after the Transformer Encoder allows the model to capture local structural information of underwater acoustic signals, such as rapid time–frequency variations, and to form high-dimensional, discriminative graph representations.
	
	To construct and update the graph, the $K$-nearest neighbors (KNN) algorithm \cite{10.1145/1963405.1963487} is employed to measure the similarity between Transformer Encoder outputs. This provides a computationally efficient and intuitive approach for graph operations, enabling the model to capture salient local relationships within the feature space while avoiding unnecessary complexity. The similarity distance is computed using the $\mathrm{p}$-norm metric:
	\begin{equation}
		\|\mathbf{x}\|_{\mathrm{p}}=\left(\sum_{i=1}^{n}\left|\mathbf{x}_{i}\right|^{\mathrm{p}}\right)^{1/\mathrm{p}},
	\end{equation}
	where $\mathrm{p}$ is set to 2 in this study. Subsequently, for each node $v_i$, $K$ nearest neighbors $\mathcal{N}(v_i)$ are connected by directed edges $e_{ji}$ from $v_j$ to $v_i$ for all $v_j \in \mathcal{N}(v_i)$. In this way, the initial Mel-graph is defined as $\mathcal{G}_{mel}=(\mathcal{V}, \mathcal{E})$, where $\mathcal{V}=\{v_1, v_2, \cdots, v_N\}$ is the node set and $\mathcal{E}$ is the edge set.
	The outputs of the Transformer Encoder, obtained through MHSA, are regarded as Mel-graph embeddings in the UATR-GTransformer. Each embedding encodes its own Mel-frequency energy distribution while also capturing global dependencies among embeddings due to the strong global modeling capability of MHSA. Consequently, these Mel-graph embeddings serve as higher-order representations that preserve detailed time–frequency information of underwater acoustic target signals, thereby implicitly constructing a robust Mel-graph.
	
\begin{figure*}
		\centering
		\includegraphics[width=16.6cm]{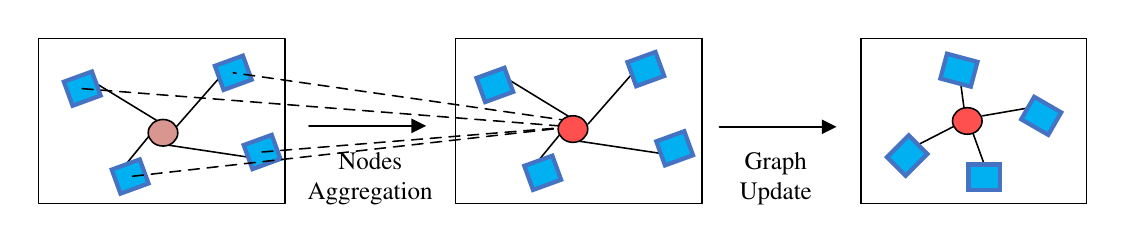}
		\caption{Illustration of graph convolution for nodes aggregation and graph update. The central node is marked by a circle, while its neighboring nodes are denoted by surrounding boxes.}
		\label{fig:gnn}
		\vspace{-1.5em}
	\end{figure*}

	The core operation of the GNN is graph convolution, which aggregates neighboring topological information and updates node features within the Mel-graph, as illustrated in Fig.~\ref{fig:gnn}.
	From the perspective of a central node $\mathbf{x}_i$, graph convolution is formulated as:
	\begin{equation}
		\mathbf{x}'_i = h(\mathbf{x}_i, g(\mathbf{x}_i, \mathcal{N}(\mathbf{x}_i); \mathbf{W}_\text{agg}); \mathbf{W}_\text{update}),
	\end{equation}
	where $g(\cdot)$ and $h(\cdot)$ denote the aggregation and update functions, respectively, and $\mathcal{N}(\mathbf{x}_i)$ is the set of neighboring nodes of $\mathbf{x}_i$. To mitigate gradient vanishing, the max-relative (MR) graph convolution \cite{deepgcn} is applied to process Mel-graph embeddings:
	\begin{equation}
		\begin{aligned}
			g(\cdot) &= \mathbf{x}_i^{\prime \prime} = \left[\mathbf{x}_i, \max \left(\left\{\mathbf{x}_j - \mathbf{x}_i \mid j \in \mathcal{N}(\mathbf{x}_i)\right\}\right)\right], \\
			h(\cdot) &= \mathbf{x}_i^{\prime} = \mathbf{x}_i^{\prime \prime} \mathbf{W}_\text{update} + \mathbf{b},
		\end{aligned}
	\end{equation}
	where $\mathbf{b}$ is the bias term. After MR graph convolution, the updated node set $\mathcal{N}(\mathbf{x}'_i)$ forms a new Mel-graph, denoted by $\mathcal{G}_{mel}'$. Here, $\mathbf{W}_\text{agg}$ and $\mathbf{W}_\text{update}$ represent learnable weights for the aggregation and update operations, respectively. In particular, the aggregation function captures salient information by computing the maximum difference between the central node and its $K$ neighbors, while the update function applies a nonlinear transformation to generate the updated graph.	
	
	After graph convolution on $\mathbf{X}$, the updated features $\mathbf{X'}$ are processed by two fully connected layers with projection matrices $\mathbf{W}_{\text{in}}$ and $\mathbf{W}_{\text{out}}$ to enhance feature diversity. A ReLU activation function is applied after the first projection layer to mitigate layer collapse. The output feature $\mathbf{Y}$ is then computed as follows:
	\begin{equation}
		\begin{gathered}
			\mathbf{X'} = \operatorname{MR\ Graph\ Convolution}(\mathbf{X}), \\
			\mathbf{Y} = \operatorname{ReLU}(\mathbf{X'} \mathbf{W}_{\text{in}})\mathbf{W}_{\text{out}} + \mathbf{X}.
		\end{gathered}
		\label{eq:gnn}
	\end{equation}

	\subsubsection{FFN}
	After GNN processing, an FFN is applied to further transform the node-level features and to integrate the Transformer and GNN modules. 
	\begin{figure}		\centering
		\includegraphics[width=8.8cm]{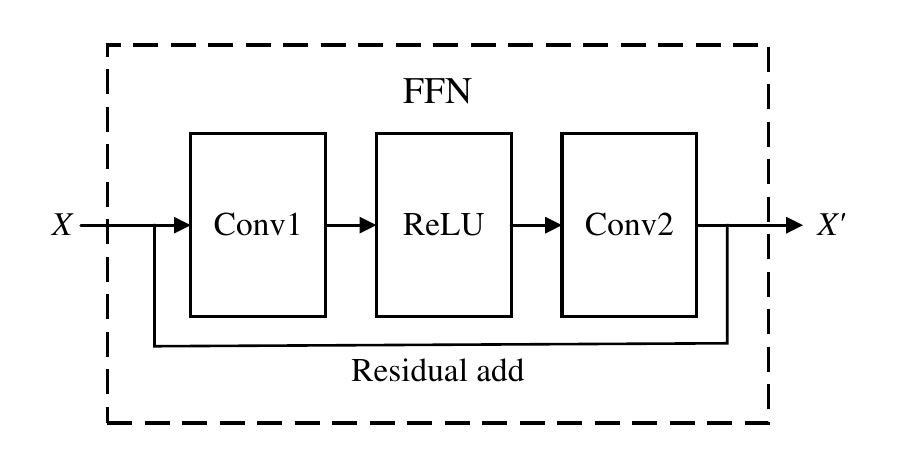}	
		\caption{Illustration of the FFN for feature transformation.}
		\label{fig:ffn}
	\end{figure}
	The structure of the FFN is illustrated in Fig.~\ref{fig:ffn} and can be expressed as:
	\begin{equation}
		\mathbf{Z} = \operatorname{ReLU}\left(\mathbf{Y}\mathbf{W}_1 + \mathbf{b}_1\right)\mathbf{W}_2 + \mathbf{b}_2 + \mathbf{Y},
	\end{equation}
	where $\mathbf{Z} \in \mathbb{R}^{N \times dim}$, $N=256$ is the number of nodes, $\mathbf{W}_1$ and $\mathbf{W}_2$ are the weights of two fully layers, and $\mathbf{b}_1$, $\mathbf{b}_2$ are the corresponding biases. The hidden dimension of the FFN is set to $4 \times dim$ to enhance its feature transformation capacity. The ReLU activation function is employed to introduce nonlinearity and improve representation learning for underwater acoustic signals.

	\subsection{Classification Head}
	To predict the ship class, a classification head is attached after the GTransformer stacks. Specifically, the classification head operates on 4-D tensors interpreted as a graph after the final FFN. Since fully connected layers alone cannot directly process such data, the classification head incorporates a pooling layer for dimension reduction and two convolutional layers to progressively extract meaningful features for prediction. 
	
	For the two convolutional layers, the first employs a $1\times1$ convolution to transform the feature map from $dim=96$ to a hidden dimension. The second $1\times1$ convolution further projects the features from the hidden dimension to $C$, where $C$ denotes the number of classes. The hidden dimension is set to 512 to better capture intricate patterns from the graph embeddings. Batch normalization and a ReLU activation are applied between the two convolutional layers to facilitate training.
	
	The overall framework of the UATR-GTransformer is summarized as follows.
	\begin{algorithm}
		\caption{UATR-GTransformer Algorithm for UATR.}  	
		\label{al:uatr_gt}
		\begin{algorithmic}[H]  
			\REQUIRE Mel-graph $x \in \mathbb{R}^{t \times f}$
			\ENSURE Classification loss $L_{ce}$\\
			\STATE 1: Apply Mel patchify on spectrogram $x$ using stem convolutions to obtain the patch set.
			\STATE 2: Add positional embedding to the patch embeddings using (\ref{eq:abs-pos}).
			\FOR{$l=1$ to $L$}
			\STATE 3: Transformer Encoder to extract deep features as Mel-graph embeddings. 
			\STATE 4: Construct Mel-graph $\mathcal{G}_{mel}=(\mathcal{V}, \mathcal{E})$ by finding $K$ nearest neighbors using the KNN algorithm. 
			\STATE 5: Graph convolution in a GNN block to aggregate information and update $\mathcal{G}_{mel}$, yielding $\mathcal{G}_{mel}'$.
			\STATE 6: FFN for feature transformation on $\mathcal{G}_{mel}'$.
			\ENDFOR
			\STATE 7: Classification head to predict the ship label $y_{\text{predict}}$.
			\STATE 8: Compute the cross-entropy loss $L_{ce}$ with the ground-truth label $y_\text{true}$.
		\end{algorithmic}
	\end{algorithm}

	\begin{table*}
		\centering
		\caption{Detailed configuration of the model architecture. The input dimension is $(B, 512, 128)$, where $B$ denotes the batch size.}
		\label{tab:UATRGT_description}
		\begin{tabular}{c|c|c|c} 
			\toprule
			{Module}                                                                             & \multicolumn{2}{>{\centering\hspace{0pt}}m{0.512\linewidth}|}{Main Opearation}               & {Dimension}  \\ \hline
			\multirow{5}{0.237\linewidth}{\hspace{0pt}\centering\arraybackslash{}Mel Patchify}               & \multicolumn{2}{>{\centering\hspace{0pt}}m{0.512\linewidth}|}{Conv(K=3, C=12, S=2, P=1)}            & (B, 12, 256, 64)    \\ 
			\cline{2-4}
			& \multicolumn{2}{>{\centering\hspace{0pt}}m{0.512\linewidth}|}{Conv(K=3, C=24, S=2, P=1)}                  & (B, 24, 128, 32)    \\ 
			\cline{2-4}
			& \multicolumn{2}{>{\centering\hspace{0pt}}m{0.512\linewidth}|}{Conv(K=3, C=48, S=2, P=1)}                  & (B, 48, 64, 16)     \\ 
			\cline{2-4}
			& \multicolumn{2}{>{\centering\hspace{0pt}}m{0.512\linewidth}|}{Conv(K=3, C=96, S=2, P=1)}                  & (B, 96, 32, 8)      \\ 
			\cline{2-4}
			& \multicolumn{2}{>{\centering\hspace{0pt}}m{0.512\linewidth}|}{Conv(K=3, C=96, S=1, P=1)}                  & (B, 96, 32, 8)      \\ 
			\hline
			\multirow{6}{0.225\linewidth}{\centering\hspace{0pt}\begin{tabular}[c]{@{}c@{}}GTransformer\\($L$=8)\end{tabular}} & Encoder                                                            & $H$=8, $dim$=96           & (B, 256, 96)        \\ 
			\cline{2-4}
			& \multirow{3}{0.231\linewidth}{\hspace{0pt}\centering\arraybackslash{}GNN} & 1$\times$1 Conv              & (B, 96, 32, 8)     \\ 
			\cline{3-4}
			&                                                                            & Graph Conv, KNN[2, 8]        & (B, 96, 256)      \\ 
			\cline{3-4}
			&                                                                            & 1$\times$1 Conv             & (B, 96, 32, 8)     \\ 
			\cline{2-4}
			& \multirow{2}{0.231\linewidth}{\hspace{0pt}\centering\arraybackslash{}FFN}  & Conv(96, 384), ReLU       & (B, 384, 32, 8)    \\ 
			\cline{3-4}
			&                                                                            &Conv(386, 96), residual connection & (B, 96, 32, 8)     \\ 
			\hline
			\multirow{3}{0.237\linewidth}{\hspace{0pt}\centering\arraybackslash{}Classification Head}            & \multicolumn{2}{>{\centering\hspace{0pt}}m{0.512\linewidth}|}{2d pooling}                         & (B, 96, 1, 1)       \\ 
			\cline{2-4}
			& \multicolumn{2}{>{\centering\hspace{0pt}}m{0.512\linewidth}|}{1$\times$1 Conv(96, 512)} & (B, 512, 1, 1)       \\ 
			\cline{2-4}
			& \multicolumn{2}{>{\centering\hspace{0pt}}m{0.512\linewidth}|}{1$\times$1 Conv(512, $C$)} & (B, $C$)              \\ \bottomrule
		\end{tabular}
	\end{table*}
	
	\section{Experimental settings}
	\subsection{Dataset description}
	The dataset used in the experiments consists of two widely researched datasets: (1) ShipsEar \cite{santos2016Shipsear}: this dataset contains a diverse collection of 90 ship audio recordings at a sampleing frequency of 52734 Hz, the duration of each recording is between 15 seconds to 10 minutes. ShipsEar contains a total of 11 vessel types, which can be further combined into 4 vessel categories depending on vessel size, and 1 background noise category. (2) DeepShip \cite{irfan2021Deepship}: this dataset consists of 265 real underwater sound recordings at a sampling frequency of 32000 Hz, which is further merged into four categories of ship vessels with no background noise provided.
	
	For preprocessing, the waveform data is first resampled to 16 kHz and then cut into 5-seconds segments. These segments are divided into training, validation, and testing sets according to time periods, using a ratio of 70\% for training, 15\% for validation, and the remainder for testing. This partitioning strategy, recommended in \cite{Niu2023}, helps prevent potential data leakage that may occur with random splitting. The detailed dataset partitions are shown in Table~\ref{dataset}.
	\begin{table*}
		\centering
		\caption{Dataset partitions of the two underwater acoustic databases.}
		\label{dataset}	
		\begin{tabular}{ccc}
			\toprule
			Dataset & \multicolumn{1}{>{\centering\hspace{0pt}}m{0.6\linewidth}}{Class} & Split sample \\ \hline
			\multirow{5}{0.2\linewidth}{\hspace{0pt}\centering\arraybackslash{}ShipsEar} & A: Fish boats, Trawlers, Mussel boat, Tugboat, Dredger & 340 \\
			& B: Motorboat, Pilotboat, Sailboat & 301 \\
			& C: Passengers & 843 \\
			& D: Ocean liner, RORO & 486 \\
			& E: Background noise & 253 \\ \hline
			\multirow{4}{0.2\linewidth}{\hspace{0pt}\centering\arraybackslash{}DeepShip} & A: Cargo & 7369 \\
			& B: Passengers & 9677 \\
			& C: Tanker & 8817 \\
			& D: Tug & 8159 \\ \bottomrule
		\end{tabular}
	\end{table*}

	\subsection{Experimental Details}
	The experiments were implemented in PyTorch (version 1.8.0) with Python (version 3.8). The hardware platform consisted of four Nvidia GeForce RTX 3090 GPUs and two Intel Xeon Platinum 8377c CPUs.  
	For data augmentation, the time–frequency masking method \cite{park2019specaugment} was applied, with a frequency mask of 24 and a time mask of 96 on the Mel-spectrogram. To ensure consistent scaling across the dataset, the input Mel-spectrograms were normalized to have zero mean and unit variance. The cross-entropy loss $L_{ce}$, a widely used loss function in recognition and classification tasks, was adopted to optimize the training process.
	
	For the training configurations, the initial learning rate was set to $1.5 \times 10^{-3}$ for ShipsEar and $1.2 \times 10^{-3}$ for DeepShip. The learning rate was decayed by a factor of 0.5 after 90 epochs for ShipsEar and 130 epochs for DeepShip. The batch size was set to 16 for ShipsEar and 64 for DeepShip, while the total number of epochs was 130 and 180, respectively. Other hyperparameters were kept the same for both datasets: the number of GTransformer blocks $L=8$; the number of nearest neighbors $K$ increased from 2 to 8 across blocks; the number of attention heads $H=8$; and the graph embedding dimension $dim=96$. These hyperparameters were determined through repeated trials to optimize recognition performance. The Adam optimizer was used to update network parameters.
	
	\subsection{Evaluation Criteria}
	The recognition performance of the proposed model was evaluated using four widely adopted metrics: overall accuracy ($OA$), average accuracy ($AA$), Kappa coefficient ($Kappa$), and $F1$-score ($F1$), averaged over five runs. Specifically, $OA$ measures overall classification accuracy, while $AA$ and $Kappa$ account for imbalanced datasets. The $F1$-score reflects the trade-off between recall and precision. Let $TP$, $TN$, $FP$, and $FN$ denote true positives, true negatives, false positives, and false negatives, respectively. These metrics are defined as follows:
	\begin{equation}
		OA=\frac{TP + TN}{TP + TN + FP + FN},
	\end{equation}
	\begin{equation}
		AA=\sum_{i=1}^n \frac{TP_i + TN_i}{TP_i + TN_i + FP_i + FN_i},
	\end{equation}
	where $TP_i$, $TN_i$, $FP_i$, and $FN_i$ represent the numbers of $TP$, $TN$, $FP$, and $FN$ for the $i$-th class.
	\begin{equation}
		Kappa=\frac{P_0 - P_e}{1 - P_e},
	\end{equation}
	where $P_0$ denotes the observed agreement among raters (equal to $OA$), and $P_e$ denotes the expected agreement by chance.
	\begin{equation}
		F1=\left(\frac{2 + \tfrac{FP}{TP} + \tfrac{FN}{TP}}{2}\right)^{-1}.
	\end{equation}
	
	\section{Results and Discussions}
	\subsection{Comparison with Baseline Models}
	To evaluate the effectiveness of the proposed UATR-GTransformer, its recognition performance is compared with other baseline DL models, including ResNet-18, DenseNet-169 \cite{sun2022underwater}, MbNet-V2 \cite{hsiao2021efficient}, Xception \cite{8099678}, EfficientNet-B0, UATR-Transformer \cite{feng2022transformer}, STM \cite{jmse10101428}, and convolution-based mixture of experts (CMoE) \cite{XIE2024123431}. The main characteristics of these baseline models are summarized below:
	
	\begin{itemize}
		\item \textbf{ResNet-18}: A residual network with 18 convolutional layers, which has demonstrated strong performance across various recognition tasks.
		
		\item \textbf{DenseNet-169}: A densely connected convolutional network with 169 layers, where each layer is connected to all preceding layers, enabling efficient feature reuse and robust recognition performance in UATR.
		
		\item \textbf{MbNet-V2}: A lightweight model based on depthwise separable convolution, which substantially reduces model parameters and computational cost while maintaining accuracy.
		
		\item \textbf{Xception}: An efficient model that also employs depthwise separable convolution, further reducing parameter count and computation without sacrificing performance. 
		
		\item \textbf{EfficientNet-B0}: An optimized model that incorporates inverted residual connections and compound scaling strategies, achieving excellent recognition accuracy with relatively low complexity. 
		
		\item \textbf{UATR-Transformer}: A convolution-free model designed to exploit both global and local information from time–frequency spectrograms for UATR tasks.
		
		\item \textbf{STM}: A Transformer-based model inspired by the Audio Spectrogram Transformer (AST) \cite{gong21b_interspeech}, specifically adapted for UATR. 
		
		\item \textbf{CMoE}: A convolutional mixture-of-experts model that adopts ResNet as its backbone to enhance feature extraction. 
	\end{itemize}
	
	\begin{table*}[h]
		\centering
		\caption{Recognition performance comparison with different methods.}
		\label{tab:comparison}
		\begin{tabular}{cccccc}
			\toprule
			Dataset                                                                        & Method            & $OA$             & $AA$             & $Kappa$          & $F1$              \\ 
			\bottomrule
			\multirow{9}{0.163\linewidth}{\hspace{0pt}\centering\arraybackslash{}ShipsEar} & ResNet-18          & 0.799          & 0.736          & 0.727          & 0.738           \\
			& DenseNet-169       & 0.798          & 0.736          & 0.726          & 0.743           \\
			& MbNet-V2          & 0.745          & 0.681          & 0.656          & 0.686           \\
			& Xception          & 0.777          & 0.765          & 0.705          & 0.766           \\
			& EfficientNet-B0   & 0.757          & 0.749          & 0.678          & 0.749           \\
			& UATR-Transformer  & 0.816          & 0.802          & 0.755          & 0.814           \\
			& STM               & 0.707          & 0.684          & 0.607          & 0.692          \\
			& CMoE        & 0.815               & 
			0.807          & 0.756         &  0.809         \\
			& UATR-GTransformer & \textbf{0.832} & \textbf{0.825} & \textbf{0.778} & \textbf{0.828}  \\ 
			\hline
			\multirow{9}{0.163\linewidth}{\hspace{0pt}\centering\arraybackslash{}DeepShip} & ResNet-18          & 0.802          & 0.796          & 0.734          & 0.799           \\
			& DenseNet-169       & 0.799          & 0.792          & 0.730          & 0.795           \\
			& MbNet-V2          & 0.630          & 0.638          & 0.509          & 0.628           \\
			& Xception          & 0.801          & 0.796          & 0.732          & 0.798           \\
			& EfficientNet-B0   & 0.795          & 0.793          & 0.725          & 0.793           \\
			& UATR-Transformer  & 0.811          & 0.806          & 0.746          & 0.808           \\
			& STM               &0.744           & 0.737         & 0.656           & 0.739           \\
			& CMoE              & 0.812               & 
			0.805          & 0.747         &  0.808         \\
			& UATR-GTransformer & \textbf{0.827} & \textbf{0.824} & \textbf{0.768} & \textbf{0.826}  \\ \bottomrule
	\end{tabular}\end{table*}
	
	To ensure fair comparisons, all networks were modified to accept 1-D Mel-spectrograms as input. Moreover, to maintain a consistent training paradigm, the SPM model was not pre-trained on ImageNet but was trained from scratch, similar to the other models.
	
	From Table~\ref{tab:comparison}, it can be observed that on the ShipsEar dataset, the proposed UATR-GTransformer achieves the best performance, with $OA=0.832$, $AA=0.825$, $Kappa=0.778$, and $F1=0.828$. On the DeepShip dataset, the UATR-GTransformer also achieves the best results, with $OA=0.827$, $AA=0.824$, $Kappa=0.768$, and $F1=0.826$. These results clearly demonstrate the effectiveness and robustness of the proposed model. Specifically, for the ShipsEar dataset, CMoE achieves the strongest performance among CNN-based methods, benefitting from its multiple expert layers that act as independent learners capable of capturing high-level patterns in underwater acoustic targets. ResNet-18 and DenseNet-169 also show competitive performance, outperforming other backbone CNNs. In contrast, the lightweight MbNet-V2, as well as EfficientNet-EfficientNet-B0, exhibit weaker performance on ShipsEar, suggesting that their relatively shallow architectures may limit the extraction of sufficiently discriminative higher-order features. Among Transformer-based approaches, the UATR-Transformer achieves moderate recognition accuracy by leveraging hierarchical tokenization and the Transformer Encoder to capture both local and global dependencies. However, STM relies on a standard square tokenization scheme, which restricts local information interaction between tokens. The lack of ImageNet pre-training further amplifies this limitation, resulting in weaker performance. On the larger DeepShip dataset, ResNet-18 and DenseNet-169 continue to demonstrate strong generalization ability, with overall accuracy values close to 0.8. Among CNNs, CMoE again achieves the best results, confirming its capability to generalize across diverse data distributions through its mixture-of-experts mechanism. Furthermore, the UATR-Transformer achieves superior performance compared to STM, demonstrating the effectiveness of its design for modeling complex underwater acoustic signals. When trained on larger datasets, both Xception and EfficientNet-B0 exhibit improved recognition accuracy, implying that increased data volumes partially offset their architectural constraints.
	
	\subsection{Ablation Study}	
	This section presents the results of ablation experiments conducted to evaluate the contribution of different components in the proposed UATR-GTransformer. In particular, we analyze the effect of the modules within the GTransformer block and the positional embedding on recognition performance, measured by the four evaluation metrics.
	
	The first set of experiments examines the importance of each module in the GTransformer block. Table~\ref{tab:ablation_1} summarizes the results obtained by removing individual components. The symbol ``–'' denotes the removal of the corresponding module. Specifically, ``– Encoder'' indicates that the model employs only the GNN and FFN in the GTransformer block, excluding the MHSA-based feature extractor. ``– GNN'' indicates that the model consists of the Encoder and FFN, but without graph embedding operations. Finally, ``– FFN'' represents the variant where the Encoder and GNN are retained, while the FFN is removed.

	\begin{table*}[h]
		\centering
		\caption{Ablation study on the GTransformer block based on the two datasets.}
		\label{tab:ablation_1}	
		\begin{tabular}{cccccc} \toprule
			Dataset      & Model             & $OA$    & $AA$    & $Kappa$ & $F1$     \\ 
			\hline
			\multirow{4}{0.167\linewidth}{\hspace{0pt}\centering\arraybackslash{}ShipsEar} & UATR-GTransformer & \textbf{0.832} & \textbf{0.825} & \textbf{0.778} & \textbf{0.828}  \\
			& ~ ~ - Encoder     & 0.780 & 0.769 & 0.709 & 0.776  \\
			& - GNN             & 0.802 & 0.800 & 0.739 & 0.801  \\
			& ~- FFN            & 0.792 & 0.783 & 0.725 & 0.788  \\ 
			\hline
			\multirow{4}{0.167\linewidth}{\hspace{0pt}\centering\arraybackslash{}DeepShip} & UATR-GTransformer & \textbf{0.827} & \textbf{0.824} & \textbf{0.768} & \textbf{0.826}  \\
			& - Encoder         & 0.818 & 0.815 & 0.756 & 0.816  \\
			& - GNN             & 0.814 & 0.811 & 0.750 & 0.812  \\
			& - FFN             & 0.815 & 0.810 & 0.751 & 0.813  \\ \bottomrule
		\end{tabular}
	\end{table*}
	\begin{table*}[h!]
		\centering
		\caption{Ablation study on the position embedding based on the two datasets.}
		\label{tab:ablation_2}
		
		\begin{tabular}{cccccc}
			\toprule
			Dataset & Model & $OA$ & $AA$ & $Kappa$ & $F1$ \\ \hline
			\multirow{3}{0.179\linewidth}{\hspace{0pt}\centering\arraybackslash{}ShipsEar} & Case 1 & 0.790 & 0.783 & 0.723 & 0.785 \\
			& Case 2 & 0.798 & 0.788 & 0.731 & 0.793 \\
			& Case 3 & \textbf{0.832} & \textbf{0.825} & \textbf{0.778} & \textbf{0.828} \\ \hline
			\multirow{3}{0.179\linewidth}{\hspace{0pt}\centering\arraybackslash{}DeepShip} & Case 1 & 0.817 & 0.817 & 0.759 & 0.818 \\
			& Case 2 & 0.821 & 0.816 & 0.760 & 0.819 \\
			& Case 3 & \textbf{0.827} & \textbf{0.824} & \textbf{0.768} & \textbf{0.826} \\ \bottomrule
		\end{tabular}
	\end{table*}
	From Table~\ref{tab:ablation_1}, it can be seen that the complete UATR-GTransformer, which incorporates the Encoder, GNN, and FFN, achieves the best $OA$, $AA$, $Kappa$, and $F1$ on both datasets. Each component within the GTransformer block contributes significantly to capturing discriminative Mel-graph representations. The Transformer Encoder, GNN, and FFN operate jointly to enhance recognition performance, and the removal of any individual component undermines the underlying Mel-graph structure, leading to noticeable performance degradation. In particular, for the ShipsEar dataset, removing any module results in substantial variation, highlighting the critical role of graph-structured feature extraction and processing for this dataset. 
	
	The second set of experiments investigates the effectiveness of the two-dimensional positional embedding $PE$ in the UATR-GTransformer. Specifically, recognition performance was compared across three configurations: Case 1, without $PE$; Case 2, with one-dimensional absolute $PE$ following standard Transformer models \cite{vaswani2017attention}; and Case 3, with two-dimensional $PE$. As shown in Table~\ref{tab:ablation_2}, introducing $PE$ consistently improves performance over Case 1, confirming its ability to capture the positional information of split patches. Moreover, Case 3 outperforms Case 2, particularly on the ShipsEar dataset, demonstrating the superiority of the two-dimensional $PE$ approach, which provides richer time–frequency distribution information for Mel-graph construction. 
	
	To further examine the contribution of the Transformer layers on the recognition performance, comparative experiments were conducted using only a single Transformer layer for initial Mel-graph embedding. Table~\ref{tab:ablation_3} shows that employing the full Transformer stack in the GTransformer block yields superior results compared to a single-layer variant, indicating that successive MHSA computations enable the extraction of higher-level semantic information across graph nodes, thereby producing more discriminative Mel-graph embeddings.
	
	\begin{table*}[h!]
		\centering
		\caption{Ablation study on the Transformer configurations based on the two datasets.}
		\label{tab:ablation_3}
		\begin{tabular}{cccccc}
			\toprule
			Dataset  &Transformer& $OA$ & $AA$ & $Kappa$ & $F1$ \\ \bottomrule
			\multirow{2}{*}{ShipsEar} & First Layer & 0.790 & 0.783 & 0.723 & 0.785 \\
			& Full layer & \textbf{0.832} & \textbf{0.825} & \textbf{0.778} & \textbf{0.828} \\ \hline
			\multirow{2}{*}{DeepShip} & First Layer & 0.817 & 0.812 & 0.754 & 0.814 \\
			&  Full layer & \textbf{0.827} & \textbf{0.824} & \textbf{0.768} & \textbf{0.826} \\ \bottomrule
		\end{tabular}
	\end{table*}
	
	Finally, it is worth noting that the ablation experiments have a smaller impact on the DeepShip dataset. This can be attributed to the larger scale of the dataset, which facilitates the learning of more generalized features and reduces the model’s reliance on individual modules.	
	\subsection{Recognition Performance under Different Features}	
	The third set of experiments evaluates the recognition performance of the UATR-GTransformer using different acoustic features, including the STFT, the Mel-Frequency Cepstral Coefficients (MFCC), and the Gammatone-Frequency Cepstral Coefficients (GFCC). These features have been widely studied for UATR \cite{10012335} and are important benchmarks for assessing the effectiveness of the proposed model. The experiments were conducted on the ShipsEar dataset for simplicity.
	\begin{table}[h]
		\centering
		\caption{Performance comparison under different features.}
		\label{tab:features}
		
		\begin{tabular}{ccccc}
			\toprule
			Feature   & $OA$    & $AA$    & $Kappa$ & $F1$     \\ 
			\midrule
			STFT      & 0.609 & 0.606 & 0.491 & 0.583  \\
			GFCC      & 0.779 & 0.773 & 0.709 & 0.772  \\
			MFCC      & 0.762 & 0.758 & 0.687 & 0.758  \\
			Mel-Fbank & \textbf{0.832} & \textbf{0.825} & \textbf{0.778} & \textbf{0.828}  \\ \bottomrule
		\end{tabular} 
	\end{table}
	As shown in Table~\ref{tab:features}, the Mel-Fbank feature yields the best recognition performance across all four evaluation metrics ($OA$, $AA$, $Kappa$, and $F1$), demonstrating that Mel-graphs provide more discriminative information for the UATR-GTransformer. In contrast, cepstral coefficient-based features (GFCC and MFCC) achieve better recognition accuracy compared with STFT, while STFT performs the worst, with an $OA$ of only 0.609. This result suggests that constructing STFT-graphs may not effectively capture discriminative information for UATR.
	
	In particular, when using the Mel-Fbank feature, the UATR-GTransformer achieves its best results on the ShipsEar dataset, with $OA=0.832$, $AA=0.825$, $Kappa=0.778$, and $F1=0.828$. Based on these findings, the Mel-Fbank feature was selected for graph embedding in the proposed UATR-GTransformer.
	
	\subsection{Parameter sensitivities}
	As major parameters of the UATR-GTransformer, we further analyze the sensitivity of $K$ in the KNN algorithm, the number of GNN blocks $L$, and the graph embedding dimension $dim$ on recognition performance using the ShipsEar dataset for simplicity. 
	
	%

	\begin{table}
		\centering
		\caption{Performance comparison under various $K$.}\label{tab:knn}
		\begin{tabular}{ccccc}
			\toprule
			$K$ & $OA$ & $AA$ & $Kappa$ & $F1$ \\ \midrule
			2 & 0.767 & 0.760 & 0.692 & 0.756 \\
			4 & 0.788 & 0.786 & 0.721 & 0.781 \\
			6 & 0.802 & 0.794 & 0.738 & 0.796 \\
			8 & 0.812 & 0.804 & 0.751 & 0.808 \\
			10 & 0.782 & 0.778 & 0.711 & 0.776 \\
			4 to 8 & 0.804 & 0.797 & 0.740 & 0.799 \\
			2 to 8 & \textbf{0.832} & \textbf{0.825} & \textbf{0.778} & \textbf{0.828} \\ \bottomrule
		\end{tabular}
	\end{table}
	
	\begin{table}
		\centering
		\caption{Recognition performance under various $L$.}\label{tab:layers}
		\begin{tabular}{ccccc}	\toprule
			$L$ & $OA$ & $AA$ & $Kappa$ & $F1$ \\ \midrule
			4 & 0.796 & 0.795 & 0.731 & 0.796 \\
			6 & 0.810 & 0.803 & 0.750 & 0.804 \\
			8 & \textbf{0.832} & \textbf{0.825} & \textbf{0.778} & \textbf{0.828} \\
			10 & 0.784 & 0.776 & 0.714 & 0.779 \\
			12 & 0.797 & 0.789 & 0.731 & 0.792  \\ \bottomrule
		\end{tabular}
	\end{table}
	\begin{table}
		\centering
		\caption{Recognition performance under various $dim$.}\label{tab:dim}
		\begin{tabular}{ccccc}	\toprule
			$dim$ & $OA$ & $AA$ & $Kappa$ & $F1$ \\ \midrule
			48 & 0.783 & 0.778 & 0.713 & 0.778 \\
			96 & \textbf{0.832} & \textbf{0.825} & \textbf{0.778} & \textbf{0.828} \\
			192 & 0.690 & 0.679 & 0.589 & 0.673 \\
			384 & 0.525 & 0.486 & 0.353 & 0.450 \\
			768 & 0.417 & 0.333 & 0.165 & 0.291 \\ \bottomrule
		\end{tabular}
	\end{table}
	\subsection{Parameter Sensitivities}
	
	Table~\ref{tab:knn} presents the recognition performance with different values of $K$ to find neighboring nodes. ``4 to 8'' indicates that $K$ is progressively increased from 4 to 8 across the GTransformer blocks. For fixed values of $K$, the best performance is obtained at $K=8$. This may be explained by the fact that splitting the Mel-spectrogram into eight frequency regions provides sufficient information for aggregating neighborhood features, whereas further increasing $K$ to 10 introduces redundancy that can reduce performance. When $K$ is gradually increased with network depth, the receptive field of the Mel-graph is enlarged, enabling information exchange among more distant nodes. This strategy is particularly beneficial for complex ship-radiated noise, as it allows the model to capture long-range dependencies and improve node separability. As shown in Table~\ref{tab:knn}, progressively enlarging $K$ improves recognition performance. In particular, the ``2 to 8'' strategy outperforms ``4 to 8'', which may be attributed to the initial layers capture local node relationships, while later layers gradually expand the receptive field and stabilize the graph structure.
	
	The number of GNN blocks $L$ and the embedding dimension $dim$ also strongly influence the generalization ability of the UATR-GTransformer, as they control the model’s depth and width. Table~\ref{tab:layers} and Table~\ref{tab:dim} report the corresponding results. From Table~\ref{tab:layers}, the optimal performance is achieved at $L=8$, suggesting that too few GNNs limit information exchange, while too many can lead to overfitting. With respect to $dim$, Table~\ref{tab:dim} shows that the best results occur at $dim=96$. A smaller $dim$ cannot adequately represent graph features, while an excessively large $dim$ produces an over-parameterized model prone to overfitting. This effect is particularly evident at $dim=768$, where $OA$ decreases sharply to 0.417.
	
	Considering these results, the following parameters are adopted for the UATR-GTransformer: $K$ increases from 2 to 8 across layers, the number of GTransformer blocks $L$ is set to 8, and the graph embedding dimension $dim$ is set to 96.


	\subsection{Statistical significance test}
	From the results in previous subsection, it is known that the UATR-GTransformer exceeds previous methods in accuracy. To quantitatively validate whether the accuracy advantages are statistically reliable, a comprehensive analysis is conducted using paired-sample t-tests, which are specifically designed for comparing paired measurements obtained under identical experimental conditions \cite{xu2017differences}. The paired-sample t-tests is particularly suitable for our evaluation framework, which utilizes the same data partitions across multiple independent runs, thereby effectively controlling for inter-run variability through its focus on within-trial performance differences.
	\begin{table*}
		\centering
		\caption{P-values of significance tests against the UATR-GTransformer.}
		\resizebox{\textwidth}{!}{
			\begin{tabular}{ccccccccc} 
				\toprule
				\multicolumn{1}{l}{} & ResNet-18 & DenseNet-169 & MbNet-V2 & Xception & EfficientNet-B0 & UATR-Transformer & STM     & CMoE     \\ 
				\hline
				ShipsEar             & $1.01 \times 10^{-2}$ & $2.30 \times 10^{-3}$ & $3.89 \times 10^{-3}$ & $3.45 \times 10^{-3}$ & $1.29 \times 10^{-5}$ & 0.149 & $1.45 \times 10^{-3}$ & $5.78 \times 10^{-2}$ \\
				DeepShip             & $4.79 \times 10^{-4}$ & $4.95 \times 10^{-3}$ & $2.40 \times 10^{-5}$ & $1.23 \times 10^{-3}$ & $6.08 \times 10^{-4}$ & $2.30 \times 10^{-4}$ & $2.46 \times 10^{-4}$ & $8.61 \times 10^{-4}$ \\
				\bottomrule
		\end{tabular}}
		\label{tab:sig}
	\end{table*}
	
	All models are evaluated using the same data splits over five repeated runs, generating paired samples for analysis. The null hypothesis for each test is a zero mean difference in $OA$. Here, we use standard significance thresholds ($p <$ 0.05 for significance, $p <$ 0.01 for strong significance). Table \ref{tab:sig} demonstrates that the proposed UATR-GTransformer achieves statistically significant improvements over most models on the ShipsEar dataset. However, because the UATR-Transformer and CMoE also deliver competitive results, the improvement over these specific models is not statistically significant. Besides, the results obtained on the DeepShip dataset provide stronger evidence, with the UATR-GTransformer achieving highly significant results against other models. 
	
	\subsection{Model Complexity}
	To further examine the computational complexity of the UATR-GTransformer, Table~\ref{tab:uatrgt_comlpex} presents comparisons on widely used complexity metrics, including the number of parameters (NP), average prediction time for a single acoustic signal (Avg.~time), giga floating-point operations (GFLOPs), and frames per second (FPS).
	\begin{table}[htbp]
		\centering
		\large
		\caption{Comparison of model complexity.}
		\label{tab:uatrgt_comlpex}
		\resizebox{\linewidth}{!}{
			\begin{tabular}{p{4cm}<{\centering}p{2.5cm}<{\centering}p{2.5cm}<{\centering}p{2.5cm}<{\centering}p{2.5cm}<{\centering}}
				\toprule
				Model & NP(M) & Avg.time(ms) & GFLOPs & FPS \\\midrule
				MbNet-V2 & 2.23 & 4.91$_{\pm 0.59}$ & 0.43 & 203.76 \\
				Xception & 3.63 & 1.82$_{\pm 0.28}$ & 0.575 & 548.18 \\
				EfficientNet-B0 & 4.01 & 9.53$_{\pm 0.63}$ & 0.54 & 104.96 \\
				ResNet-18 & 11.17 & 3.24$_{\pm 0.57}$ & 2.28 & 309.15 \\
				DenseNet-169 & 12.49 & 42.54$_{\pm 5.99}$ & 4.41 & 23.51 \\
				UATR-Transformer & 2.55 & 3.54$_{\pm 0.43}$ & 3.25 & 282.95 \\
				CMoE & 11.19 & 4.28$_{\pm 0.49}$ & 2.28 & 233.47 \\
				UATR-GTransformer & 2.05 & 18.99$_{\pm 0.72}$ & 0.672 & 52.65 \\
				\bottomrule
		\end{tabular}}
	\end{table}
	
	As shown in Table~\ref{tab:uatrgt_comlpex}, the UATR-GTransformer has a relatively small NP and low GFLOPs, but exhibits higher Avg.~time and lower FPS compared with most other models. This is likely due to the additional computations required for similarity calculations and multi-head self-attention across multiple nodes. Among lightweight CNNs, MbNet-V2, Xception, and EfficientNet-B0 all show low GFLOPs, indicating less computational requirements. Owing to its larger spatial resolution and wider network width, EfficientNet-B0 contains the largest number of parameters (4.01M) among lightweight CNNs and yields the slowest prediction speed, with an Avg.~time of 9.53 ms. In contrast, Xception achieves the fastest prediction owing to the use of depthwise and pointwise convolutions, and also has the smallest NP and GFLOPs, thereby demonstrating the best recognition efficiency. For ResNet-based models, CMoE provides higher recognition performance than ResNet-18, though with slightly greater complexity, which may be attributed to the introduction of the mixture-of-experts mechanism. DenseNet-169, due to its dense connections within a deep architecture, exhibits the highest complexity overall, with 12.49M parameters, an Avg.~time of 42.54$_{\pm 5.99}$ ms, GFLOPs of 4.41, and the lowest FPS (23.51).
	
	\subsection{Interpretability experiments}
\begin{figure*}[h]
		\centering
		\includegraphics[width=16.8cm]{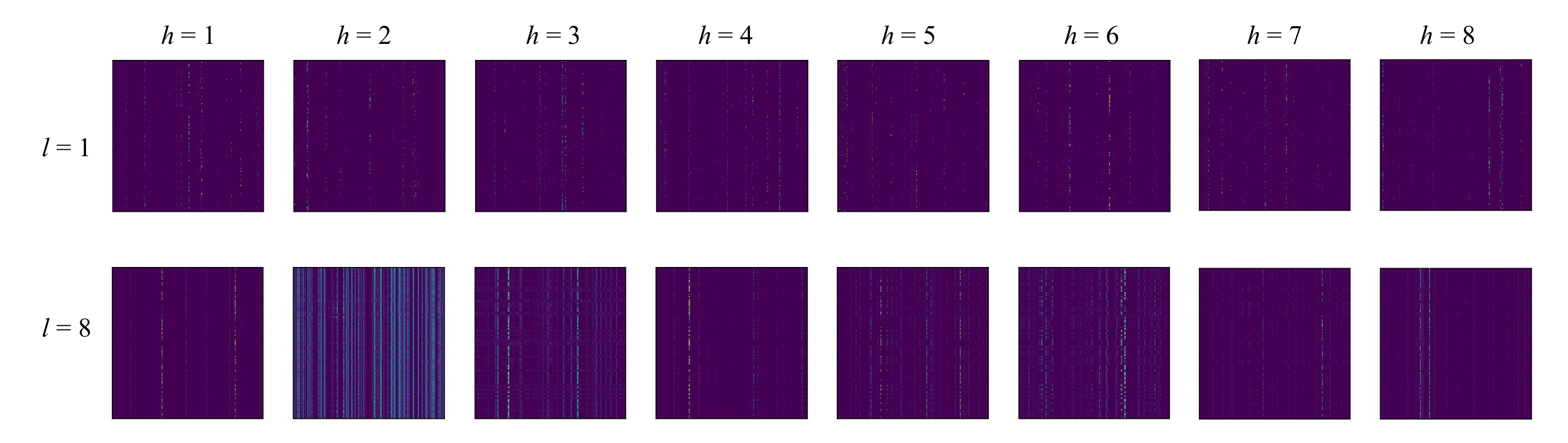}
		\caption{Visualization of attention matrices in the first and last Transformer Encoder layers using Mel-spectrogram features. $l \in [1,8]$ denotes the $l$-th GTransformer Block, and $h \in [1,8]$ the $h$-th attention head.}
		\label{fig:heads_uatrgt}
	\end{figure*}

	In the UATR-GTransformer, information flows through the Transformer Encoder via the attention matrix, which enables the model to capture dependencies among Mel-graph embeddings from split spectrogram patches. To investigate how attention operates, we first visualize the attention matrices from the $H=8$ attention heads in the UATR-GTransformer. Fig.~\ref{fig:heads_uatrgt} shows the $256 \times 256$ attention matrices from the eight heads in the first and last Transformer Encoder layers when a Mel-spectrogram is processed. The horizontal and vertical axes represent the positions of queries and keys, respectively, and the values indicate their similarity. The presence of vertical line patterns suggests that a query attends to multiple keys, reflecting the model’s capacity to perceive global structures and capture high-level information through multi-head interactions.
	
	As shown in Fig.~\ref{fig:heads_uatrgt}, the first-layer attention heads display relatively sparse vertical line patterns, indicating that they primarily capture localized embedding details with limited importance. By contrast, in the final layer, the attention becomes more concentrated on multiple embeddings, with stronger interactions among nodes. For example, the second attention head ($h=2$) highlights several prominent vertical lines, demonstrating that important information is aggregated across multiple embeddings. These results confirm that stacking GTransformer blocks progressively enhances global feature perception, enabling the model to capture higher-order information from the Mel-spectrogram.
	
	\begin{figure*}[h]
		\centering
		\includegraphics[width=15.6cm]{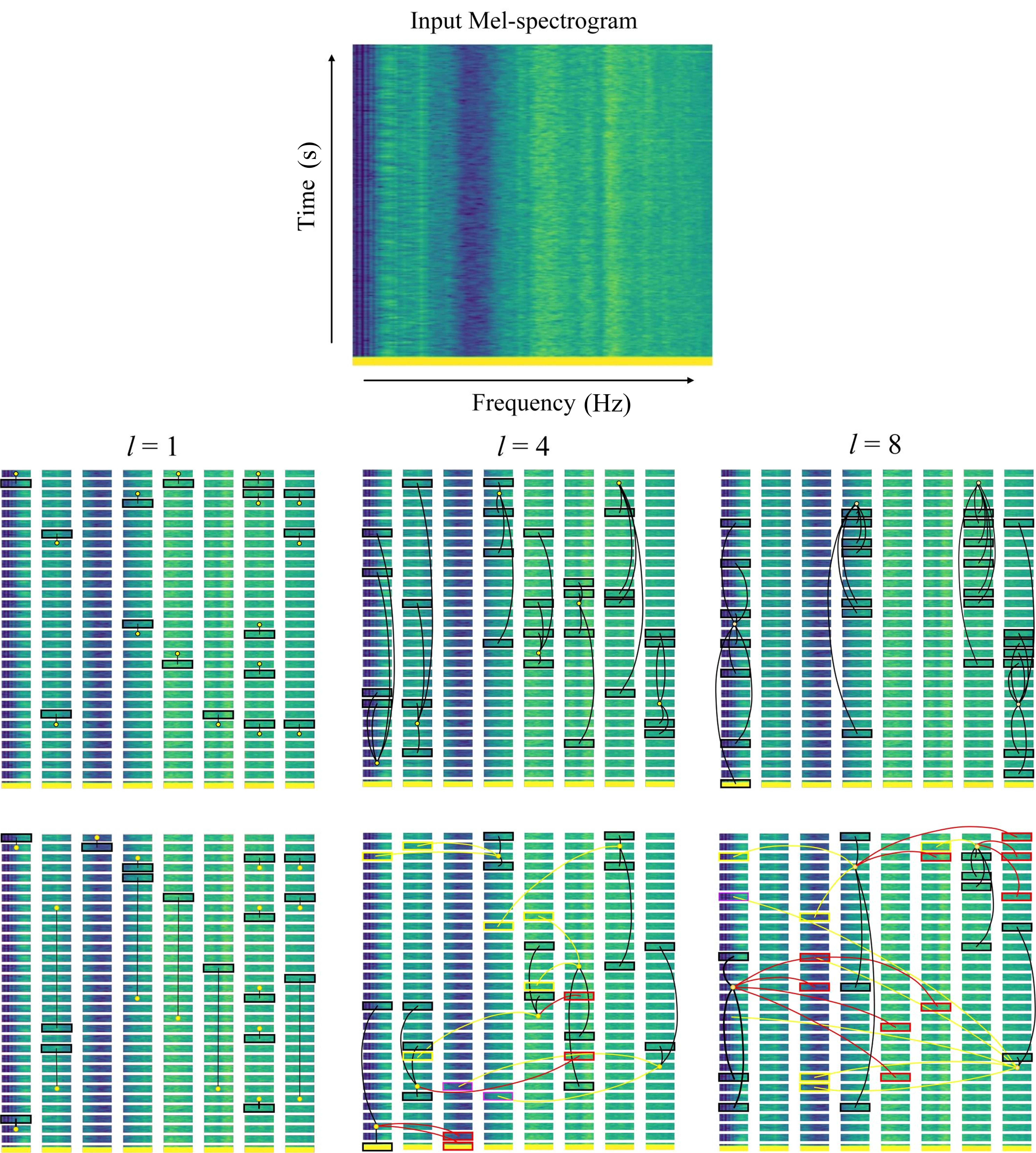}
		\caption{Visualization of Mel-graph connections for an input Mel-spectrogram. The central node is shown as a circle, while neighboring nodes are shown as surrounding boxes. Row 1: graph visualization without the Transformer Encoder (only GNN). Row 2: graph visualization with the complete UATR-GTransformer. $l$ denotes the $l$-th GTransformer Block.}
		\label{fig:graph_visualization}
		\vspace{-1em}
	\end{figure*}
	
	To further examine graph structure learning, the learned Mel-graph is visualized in Fig.~\ref{fig:graph_visualization}. The input Mel-spectrogram is partitioned into $32 \times 8$ patches, corresponding to 256 graph nodes. Row 1 shows the Mel-graph learned by the model without the Transformer Encoder, where only the GNN is applied. Row 2 shows the Mel-graph learned by the complete UATR-GTransformer. 
	
	In Row 1, the GNN primarily extracts frequency-domain features to build discriminative criteria. In the first block ($l=1$), neighboring nodes are identified along the adjacent time axis. When $l=4$ with $K=4$, neighbors are primarily within the same frequency bands. At the final block ($l=8$), with $K=8$, the receptive field expands, allowing broader frequency-domain interactions. These results suggest that the Mel-graph learned by GNNs is mainly frequency-driven, with nodes in the same bands more tightly connected.
	
	Row 2 illustrates the effect of combining the Transformer Encoder with the GNN. At $l=1$, MHSA facilitates global interactions by linking adjacent time-frequency bands as well as distant frequency nodes. As $l$ increases, the receptive field expands further. At $l=4$, the model begins to capture long-range relationships both within and across frequency bands. At $l=8$, the UATR-GTransformer integrates both local frequency-domain connections and global cross-band interactions, enabling a more comprehensive representation of the signal.
	
	In summary, the interpretability experiments highlight complementary roles of the Transformer Encoder and GNN. The Transformer Encoder enhances global perception across frequency bands and captures complex time–frequency relationships through MHSA, while the GNN emphasizes local frequency-domain consistency, ensuring that discriminative information is preserved.

	\section{Conclusion}
	This paper proposes an intelligent UATR approach based on a non-Euclidean framework, named as the UATR-GTransformer. In this model, the input Mel-spectrogram is first divided into overlapping patches, which are processed by a Transformer Encoder to obtain graph embeddings enriched with Mel-frequency information. These embeddings are treated as graph nodes and connected via the KNN algorithm to construct a Mel-graph that captures the topological structure of the acoustic signal. A GNN and an FFN are then employed to enhance the feature representations and perform classification, followed by a classification head for final prediction. Experimental results demonstrate that the UATR-GTransformer achieves superior performance compared with baseline models, validating its effectiveness.
	
	In contrast to conventional methods that treat spectrograms as images, the UATR-GTransformer represents time-frequency patches as nodes in a graph, enabling the capture of internal relationships between features and the construction of local structures through KNN graphs. The interpretability experiments further show that the UATR-GTransformer provides valuable insights into the information flow and decision-making process.
	
	Despite its contributions, several limitations remain. First, the experiments were conducted only on two publicly available datasets; thus, the model’s generalization ability to unseen sea areas and conditions requires further validation. Second, the computational complexity of the UATR-GTransformer is relatively high due to the similarity calculations and MHSA among multiple nodes, which may restrict its real-time applicability. Future work may focus on optimizing the architecture to reduce complexity and facilitate real-time deployment. Finally, while the model offers a degree of interpretability by illustrating local feature relationships through GNNs, it does not yet provide detailed insights into the most critical frequency bands. Further research will therefore explore graph feature quantification techniques with higher-quality underwater acoustic datasets.


	\bibliography{myreference}{}

@ARTICLE {hypertools,
	author  = {Andrew C. Heusser and Kirsten Ziman and Lucy L. W. Owen and Jeremy R. Manning},    
	title   = {Hyper{T}ools: a {P}ython Toolbox for Gaining Geometric Insights into High-Dimensional Data},    
	journal = {Journal of Machine Learning Research},
	year    = {2018},
	volume  = {18},	
	number  = {152},	
	pages   = {1-6}
}

@article{xie2022adaptive,
	title={Adaptive ship-radiated noise recognition with learnable fine-grained wavelet transform},
	author={Xie, Yuan and Ren, Jiawei and Xu, Ji},
	journal={Ocean Engineering},
	volume={265},
	pages={112626},
	month=dec,
	year={2022},
	publisher={Elsevier}
}

@article{sun2022underwater,
	title={Underwater single-channel acoustic signal multitarget recognition using convolutional neural networks},
	author={Sun, Qinggang and Wang, Kejun},
	journal={The Journal of the Acoustical Society of America},
	volume={151},
	number={3},
	pages={2245--2254},
	year={2022},
	month=mar,
	publisher={AIP Publishing}
}

@article{10.1121/10.0015053,
	author = {Xie, Yuan and Ren, Jiawei and Xu, Ji},
	title = {Underwater-art: Expanding information perspectives with text templates for underwater acoustic target recognition},
	journal = {The Journal of the Acoustical Society of America},
	volume = {152},
	number = {5},
	pages = {2641-2651},
	year = {2022},
	month = nov,
	issn = {0001-4966},
	doi = {10.1121/10.0015053}
}

@article{10.1121/10.0019937,
	author = {Xu, Kele and Xu, Qisheng and You, Kang and Zhu, Boqing and Feng, Ming and Feng, Dawei and Liu, Bo},
	title = {Self-supervised learning–based underwater acoustical signal classification via mask modeling},
	journal = {The Journal of the Acoustical Society of America},
	volume = {154},
	number = {1},
	pages = {5-15},
	year = {2023},
	month = jul,
	issn = {0001-4966},
	doi = {10.1121/10.0019937}
}

@article{yang2020underwater,
	title={Underwater acoustic research trends with machine learning: passive {SONAR} applications},
	author={Yang, Haesang and Lee, Keunhwa and Choo, Youngmin and Kim, Kookhyun},
	journal={Journal of Ocean Engineering and Technology},
	volume={34},
	number={3},
	pages={227--236},
	year={2020},
	month=jun,
	publisher={Ocean Engineering and Technology}
}

@ARTICLE{doan2020underwater,
	author={Doan, Van-Sang and Huynh-The, Thien and Kim, Dong-Seong},
	journal={IEEE Geoscience and Remote Sensing Letters}, 
	title={Underwater Acoustic Target Classification Based on Dense Convolutional Neural Network}, 
	year={2022},
	month=oct,
	volume={19},
	number={},
	pages={1-5},
	doi={10.1109/LGRS.2020.3029584}}

@article{feng2022transformer,
	title={A {T}ransformer-Based Deep Learning Network for Underwater Acoustic Target Recognition},
	author={Feng, Sheng and Zhu, Xiaoqian},
	journal={IEEE Geoscience and Remote Sensing Letters},
	volume={19},
	pages={1--5},
	year={2022},
	month=aug,
	publisher={IEEE}
}

@article{esfahanian2013using,
	title={Using local binary patterns as features for classification of dolphin calls},
	author={Esfahanian, M and Zhuang, H and Erdol, N},
	journal={The Journal of the Acoustical Society of America},
	volume={134},
	number={1},
	pages={EL105--EL111},
	year={2013},
	month=jun,
	publisher={AIP Publishing}
}

@inproceedings{gong2022ssast,
	title={{SSAST}: Self-{S}upervised {A}udio {S}pectrogram {T}ransformer},
	author={Gong, Yuan and Lai, Cheng-I and Chung, Yu-An and Glass, James},
	booktitle={Proceedings of the AAAI Conference on Artificial Intelligence},
	volume={36},
	number={10},
	pages={10699--10709},
	year={2022}
}

@inproceedings{deepgcn,
	title={Deep{GCN}s: Can {GCN}s go as deep as {CNN}s?},
	author={Li, Guohao and Muller, Matthias and Thabet, Ali and Ghanem, Bernard},
	booktitle={2019 IEEE/CVF International Conference on Computer Vision (ICCV)},
	pages={9267--9276},
	year={2019},
	publisher = {IEEE Computer Society},
address = {Los Alamitos, CA, USA},
month =Nov
}

@Article{s21041429,
	AUTHOR = {Hu, Gang and Wang, Kejun and Liu, Liangliang},
	TITLE = {Underwater Acoustic Target Recognition Based on Depthwise Separable Convolution Neural Networks},
	JOURNAL = {Sensors},
	VOLUME = {21},
	YEAR = {2021},
	NUMBER = {4},
	pages = {1429},
	ISSN = {1424-8220},
	DOI = {10.3390/s21041429}
}

@inproceedings{vaswani2017attention,
	author = {Vaswani, Ashish and Shazeer, Noam and Parmar, Niki and Uszkoreit, Jakob and Jones, Llion and Gomez, Aidan N. and Kaiser, \L{}ukasz and Polosukhin, Illia},
	title = {Attention is all you need},
	year = {2017},
	isbn = {9781510860964},
	publisher = {Curran Associates Inc.},
	address = {Red Hook, NY, USA},
	booktitle = {Proceedings of the 31st International Conference on Neural Information Processing Systems},
	pages = {6000–6010},
	numpages = {11},
	location = {Long Beach, California, USA},
	series = {NIPS'17}
}

@article{xiao2021underwater,
	title={Underwater acoustic target recognition using attention-based deep neural network},
	author={Xiao, Xu and Wang, Wenbo and Ren, Qunyan and Gerstoft, Peter and Ma, Li},
	journal={JASA Express Letters},
	volume={1},
	number={10},
	pages={106001},
	year={2021},
	month=oct,
	publisher={Acoustical Society of America}
}

@inproceedings{gong21b_interspeech,
	author={Yuan Gong and Yu-An Chung and James Glass},
	title={{{AST}: {A}udio {S}pectrogram {T}ransformer}},
	year=2021,
	booktitle={Proc. Interspeech 2021},
	pages={571--575},
	doi={10.21437/Interspeech.2021-698}
}

@article{santos2016shipsear,
	title={Ships{E}ar: An underwater vessel noise database},
	author={Santos-Dom{\'\i}nguez, David and Torres-Guijarro, Soledad and Cardenal-L{\'o}pez, Antonio and Pena-Gimenez, Antonio},
	journal={Applied Acoustics},
	volume={113},
	pages={64--69},
	year={2016},
	publisher={Elsevier}
}

@article{irfan2021deepship,
	title={Deep{S}hip: An underwater acoustic benchmark dataset and a separable convolution based autoencoder for classification},
	author={Irfan, Muhammad and Jiangbin, Zheng and Ali, Shahid and Iqbal, Muhammad and Masood, Zafar and Hamid, Umar},
	journal={Expert Systems with Applications},
	volume={183},
	pages={115270},
	year={2021},
	month=nov,
	publisher={Elsevier}
}

@inproceedings{park2019specaugment,
	title={Spec{A}ugment: A simple data augmentation method for automatic speech recognition},
	author={Park, Daniel S and Chan, William and Zhang, Yu and Chiu, Chung-Cheng and Zoph, Barret and Cubuk, Ekin D and Le, Quoc V},
	  booktitle = {Proc. Interspeech 2019},
  year = {2019},
pages = {2613-2617},
month = sep
}

@article{tian2023joint,
	title={Joint learning model for underwater acoustic target recognition},
	author={Tian, Sheng-Zhao and Chen, Duan-Bing and Fu, Yan and Zhou, Jun-Lin},
	journal={Knowledge-Based Systems},
	volume={260},
	pages={110119},
	year={2023},
	month=jan,
	publisher={Elsevier}
}

@ARTICLE{10012335,
	author={Wang, Biao and Zhang, Wei and Zhu, Yunan and Wu, Chengxi and Zhang, Shizhen},
	journal={IEEE Geoscience and Remote Sensing Letters}, 
	title={An Underwater Acoustic Target Recognition Method Based on {AMN}et}, 
	year={2023},
	month=jan,
	volume={20},
	number={},
	pages={1-5},
	doi={10.1109/LGRS.2023.3235659}}

@INPROCEEDINGS{7435957,
	author={de Moura, Natanael Nunes and de Seixas, José Manoel},
	booktitle={2015 Latin America Congress on Computational Intelligence (LA-CCI)}, 
	title={Novelty detection in passive {SONAR} systems using support vector machines}, 
	year={2015},
	month=Oct,
	volume={},
	number={},
	pages={1-6},
	doi={10.1109/LA-CCI.2015.7435957}}

@INPROCEEDINGS{7108260,
	author={Sherin B. M. and Supriya M. H.},
	booktitle={2015 IEEE Underwater Technology (UT)}, 
	title={Selection and parameter optimization of {SVM} kernel function for underwater target classification}, 
	year={2015},
	month=Feb,
	volume={},
	number={},
	pages={1-5},
	doi={10.1109/UT.2015.7108260}}

@article{ZHOU2023115784,
	title = {An attention-based multi-scale convolution network for intelligent underwater acoustic signal recognition},
	journal = {Ocean Engineering},
	volume = {287},
	pages = {115784},
	year = {2023},
	month=nov,
	issn = {0029-8018},
	doi = {https://doi-org-s.libyc.nudt.edu.cn:443/10.1016/j.oceaneng.2023.115784},
	author = {Aolong Zhou and Xiaoyong Li and Wen Zhang and Chengwu Zhao and Kaijun Ren and Yanxin Ma and Junqiang Song}
}

@Article{jmse10101428,
	AUTHOR = {Li, Peng and Wu, Ji and Wang, Yongxian and Lan, Qiang and Xiao, Wenbin},
	TITLE = {{STM}: Spectrogram Transformer Model for Underwater Acoustic Target Recognition},
JOURNAL = {Journal of Marine Science and Engineering},
VOLUME = {10},
YEAR = {2022},
month=oct,
NUMBER = {10},
pages = {1428},
	ARTICLE-NUMBER = {1428},
	ISSN = {2077-1312},
	DOI = {10.3390/jmse10101428}
}

@inproceedings{DBLP:conf/interspeech/GongCG21,
	author       = {Yuan Gong and
	Yu{-}An Chung and
	James R. Glass},
	title        = {{AST:} {A}udio {S}pectrogram {T}ransformer},
	booktitle    = {Interspeech 2021, 22nd Annual Conference of the International Speech
	Communication Association},
	pages        = {571--575},
	publisher    = {{ISCA}},
	year         = {2021},
	doi          = {10.21437/INTERSPEECH.2021-698}
}

@article{YANG2024107983,
	title = {Underwater acoustic target recognition based on sub-band concatenated {M}el spectrogram and multidomain attention mechanism},
	journal = {Engineering Applications of Artificial Intelligence},
	volume = {133},
	pages = {107983},
	year = {2024},
	month=jul,
	issn = {0952-1976},
	doi = {https://doi-org-s.libyc.nudt.edu.cn:443/10.1016/j.engappai.2024.107983},
	author = {Shuang Yang and Anqi Jin and Xiangyang Zeng and Haitao Wang and Xi Hong and Menghui Lei}
}

@Article{Waikhom2023,
	author={Waikhom, Lilapati
	and Patgiri, Ripon},
	title={A survey of graph neural networks in various learning paradigms: methods, applications, and challenges},
	journal={Artificial Intelligence Review},
	year={2023},
	month=Jul,
	day={01},
	volume={56},
	number={7},
	pages={6295-6364},
	issn={1573-7462},
	doi={10.1007/s10462-022-10321-2}
}

@Article{Niu2023,
	author={Niu, Haiqiang
	and Li, Xiaolei
	and Zhang, Yonglin
	and Xu, Ji},
	title={Advances and applications of machine learning in underwater acoustics},
	journal={Intelligent Marine Technology and Systems},
	year={2023},
	month=Oct,
	day={20},
	volume={1},
	number={1},
	pages={8},
	issn={2948-1953},
	doi={10.1007/s44295-023-00005-0}
}

@inproceedings{hsiao2021efficient,
	title={Efficient Computation of Depthwise Separable Convolution in {M}oblie{N}et Deep Neural Network Models},
	author={Hsiao, Shen-Fu and Tsai, Bo-Ching},
	booktitle={2021 IEEE International Conference on Consumer Electronics-Taiwan (ICCE-TW)},
	pages={1--2},
	year={2021},
	month=sep,
	organization={IEEE}}

@INPROCEEDINGS{8099678,
	author={Chollet, François},
	booktitle={2017 IEEE Conference on Computer Vision and Pattern Recognition (CVPR)}, 
	title={Xception: Deep Learning with Depthwise Separable Convolutions}, 
	year={2017},
	month=jul,
	volume={},
	number={},
	pages={1800-1807},
	doi={10.1109/CVPR.2017.195}}

@article{liu2021underwater,
	title = {Underwater target recognition using convolutional recurrent neural networks with 3-{D} {M}el-spectrogram and data augmentation},
	journal = {Applied Acoustics},
	volume = {178},
	pages = {107989},
	year = {2021},
	month=jul,
	issn = {0003-682X},
	doi = {https://doi.org/10.1016/j.apacoust.2021.107989},
	author = {Feng Liu and Tongsheng Shen and Zailei Luo and Dexin Zhao and Shaojun Guo}
}

@article{10.1121/1.5133944,
	author = {Bianco, Michael J. and Gerstoft, Peter and Traer, James and Ozanich, Emma and Roch, Marie A. and Gannot, Sharon and Deledalle, Charles-Alban},
	title = "{Machine learning in acoustics: Theory and applications}",
	journal = {The Journal of the Acoustical Society of America},
	volume = {146},
	number = {5},
	pages = {3590-3628},
	year = {2019},
	month = nov,
	issn = {0001-4966},
	doi = {10.1121/1.5133944}
}

@ARTICLE{10414073,
	author={Feng, Sheng and Zhu, Xiaoqian and Ma, Shuqing},
	journal={IEEE/ACM Transactions on Audio, Speech, and Language Processing}, 
	title={Masking Hierarchical {T}okens for Underwater Acoustic Target Recognition With Self-Supervised Learning}, 
	year={2024},
	month=jan,
	volume={32},
	number={},
	pages={1365-1379},
	doi={10.1109/TASLP.2024.3358719}}

@INPROCEEDINGS{10390008,
	author={Quraishi, Suhail Javed and Singh, Malikhan and Prasad, Sujit Kumar and Arora, Kavita and Pathak, Sonal and Singh, Anupam},
	booktitle={2023 3rd International Conference on Technological Advancements in Computational Sciences (ICTACS)}, 
	title={A Machine Learning Approach to Rock and Mine Classification in {SONAR} Systems Using Logistic Regression}, 
	year={2023},
	month=nov,
	volume={},
	number={},
	pages={462-468},
	doi={10.1109/ICTACS59847.2023.10390008}}

@InProceedings{10.1007/978-981-99-7505-1_61,
	author="Song, Ziyi
	and Ma, Lin",
	title="Speech Command Recognition Algorithm Based on Improved {MFCC} Features",
	booktitle="Communications, Signal Processing, and Systems",
	year="2024",
	month=mar,
	publisher="Springer Nature Singapore",
	address="Singapore",
	pages="587--595",
	isbn="978-981-99-7505-1"
}

@article{Hinich1982,
	author = {Hinich, Melvin J.},
	title = {TESTING FOR {G}AUSSIANITY AND LINEARITY OF A STATIONARY TIME SERIES},
	journal = {Journal of Time Series Analysis},
	volume = {3},
	number = {3},
	pages = {169-176},
	doi = {https://doi.org/10.1111/j.1467-9892.1982.tb00339.x},
	year = {1982},
	month=may
}

@ARTICLE{7763882,
	author={Mei, Jonathan and Moura, José M. F.},
	journal={IEEE Transactions on Signal Processing}, 
	title={Signal Processing on Graphs: Causal Modeling of Unstructured Data}, 
	year={2017},
	month=dec,
	volume={65},
	number={8},
	pages={2077-2092},
	doi={10.1109/TSP.2016.2634543}}

@ARTICLE{9526764,
	author={Torkamani, Razieh and Zayyani, Hadi and Marvasti, Farokh},
	journal={IEEE Transactions on Circuits and Systems II: Express Briefs}, 
	title={Joint Topology Learning and Graph Signal Recovery Using Variational Bayes in Non-{G}aussian Noise}, 
	year={2022},
	month=Sep,
	volume={69},
	number={3},
	pages={1887-1891},
	doi={10.1109/TCSII.2021.3109339}}

@article{PhysRevE.92.022817,
	title = {Time reversibility from visibility graphs of nonstationary processes},
	author = {Lacasa, Lucas and Flanagan, Ryan},
	journal = {Phys. Rev. E},
	volume = {92},
	issue = {2},
	pages = {022817},
	numpages = {13},
	year = {2015},
	month = Aug,
	publisher = {American Physical Society},
	doi = {10.1103/PhysRevE.92.022817}
}

@article{JMLR:v9:vandermaaten08a,
	author  = {Laurens van der Maaten and Geoffrey Hinton},
	title   = {Visualizing Data using t-{SNE}},
	journal = {Journal of Machine Learning Research},
	year    = {2008},
	volume  = {9},
	number  = {86},
	pages   = {2579--2605}
}

@article{TORRES2024111268,
	title = {Multi-scale cross-attention {T}ransformer via graph embeddings for few-shot molecular property prediction},
	journal = {Applied Soft Computing},
	volume = {153},
	pages = {111268},
	year = {2024},
	month=mar,
	issn = {1568-4946},
	doi = {https://doi.org/10.1016/j.asoc.2024.111268},
	author = {Luis H.M. Torres and Bernardete Ribeiro and Joel P. Arrais}
}

@ARTICLE{10530642,
	author={Han, Peng and Zhang, Xiangliang},
	journal={International Journal of Crowd Science}, 
	title={{VGE}: {G}ene-Disease Association by Variational Graph Embedding}, 
	year={2024},
	month=may,
	volume={8},
	number={2},
	pages={95-99},
	doi={10.26599/IJCS.2024.9100004}}

@article{10.1121/10.0003382,
	author = {Thode, Aaron M. and Conrad, Alexander S. and Ozanich, Emma and King, Rylan and Freeman, Simon E. and Freeman, Lauren A. and Zgliczynski, Brian and Gerstoft, Peter and Kim, Katherine H.},
	title = "{Automated two-dimensional localization of underwater acoustic transient impulses using vector sensor image processing (vector sensor localization)}",
	journal = {The Journal of the Acoustical Society of America},
	volume = {149},
	number = {2},
	pages = {770-787},
	year = {2021},
	month = Feb,
	issn = {0001-4966},
	doi = {10.1121/10.0003382}
}

@article{10.1121/1.4776775,
	author = {Lani, Shane W. and Sabra, Karim G. and Hodgkiss, William S. and Kuperman, W. A. and Roux, Philippe},
	title = "{Coherent processing of shipping noise for ocean monitoring}",
	journal = {The Journal of the Acoustical Society of America},
	volume = {133},
	number = {2},
	pages = {EL108-EL113},
	year = {2013},
	month = jan,
	doi = {10.1121/1.4776775}
}

@inproceedings{10.1145/1963405.1963487,
	author = {Dong, Wei and Moses, Charikar and Li, Kai},
	title = {Efficient k-nearest neighbor graph construction for generic similarity measures},
	year = {2011},
	month=mar,
	isbn = {9781450306324},
	publisher = {Association for Computing Machinery},
	address = {New York, NY, USA},
	doi = {10.1145/1963405.1963487},
	pages = {577–586},
	numpages = {10},
	location = {Hyderabad, India}
}

@inproceedings{10.5555/3540261.3542586,
	author       = {Tete Xiao and
Mannat Singh and
Eric Mintun and
Trevor Darrell and
Piotr Doll{\'{a}}r and
Ross B. Girshick},
title        = {Early Convolutions Help {T}ransformers See Better},
booktitle    = {Advances in Neural Information Processing Systems 34: Annual Conference
on Neural Information Processing Systems 2021},
pages        = {30392--30400},
year         = {2021},
month=dec
}

@article{XIE2024123431,
	title = {Unraveling complex data diversity in underwater acoustic target recognition through convolution-based mixture of experts},
	journal = {Expert Systems with Applications},
	volume = {249},
	pages = {123431},
	year = {2024},
	month=sep,
	issn = {0957-4174},
	author = {Yuan Xie and Jiawei Ren and Ji Xu}
}

@misc{Swami2025,
	author = {Ananthram Swami},
	title = {{HOSA}-Higher Order Spectral Analysis Toolbox},
	  howpublished = {MATLAB Central File Exchange},
	 month = sep,
	 year = {2025},
	 note = {[Online]},
	 url = {https://www.mathworks.com/matlabcentral/fileexchange/3013-hosa-higher-order-spectral-analysis-toolbox.},
	 urldate = {2025-09-09}
}

@article{xu2017differences,
	title = {The Differences and Similarities Between Two-Sample \textit{T}-Test and Paired \textit{T}-Test},
	author = {Xu, M. and Fralick, D. and Zheng, J. Z. and Wang, B. and Tu, X. M. and Feng, C.},
	journal = {Shanghai Archives of Psychiatry},
	year = {2017},
	month=jun,
	volume = {29},
	number = {3},
	pages = {184--188},
	doi = {10.11919/j.issn.1002-0829.217070}
}
	\bibliographystyle{IEEEtran}



\vspace{-33pt}
\begin{IEEEbiography}[{\includegraphics[width=1in,height=1.25in,clip,keepaspectratio]{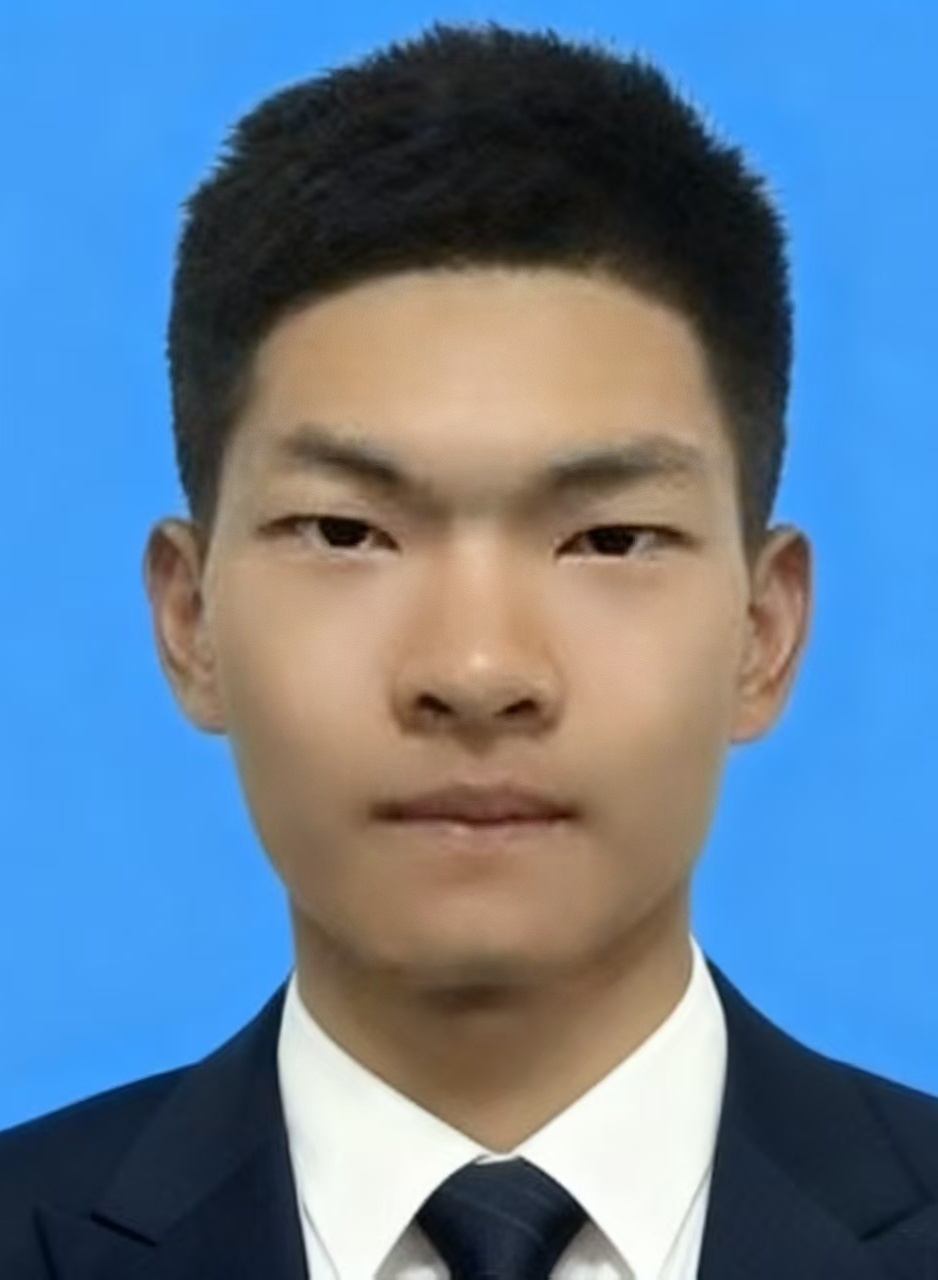}}]{Sheng Feng} received the Ph.D. degree in computer science and technology from National University of Defense Technology, Changsha, China, in 2024. He is currently an Assistant Researcher with the College of Meteorology and Oceanography, National University of Defense Technology. His research interests include ocean information processing, artificial intelligence, underwater acoustic target recognition and tracking.  
\end{IEEEbiography}

\vspace{-33pt}
\begin{IEEEbiography}[{\includegraphics[width=1in,height=1.25in,clip,keepaspectratio]{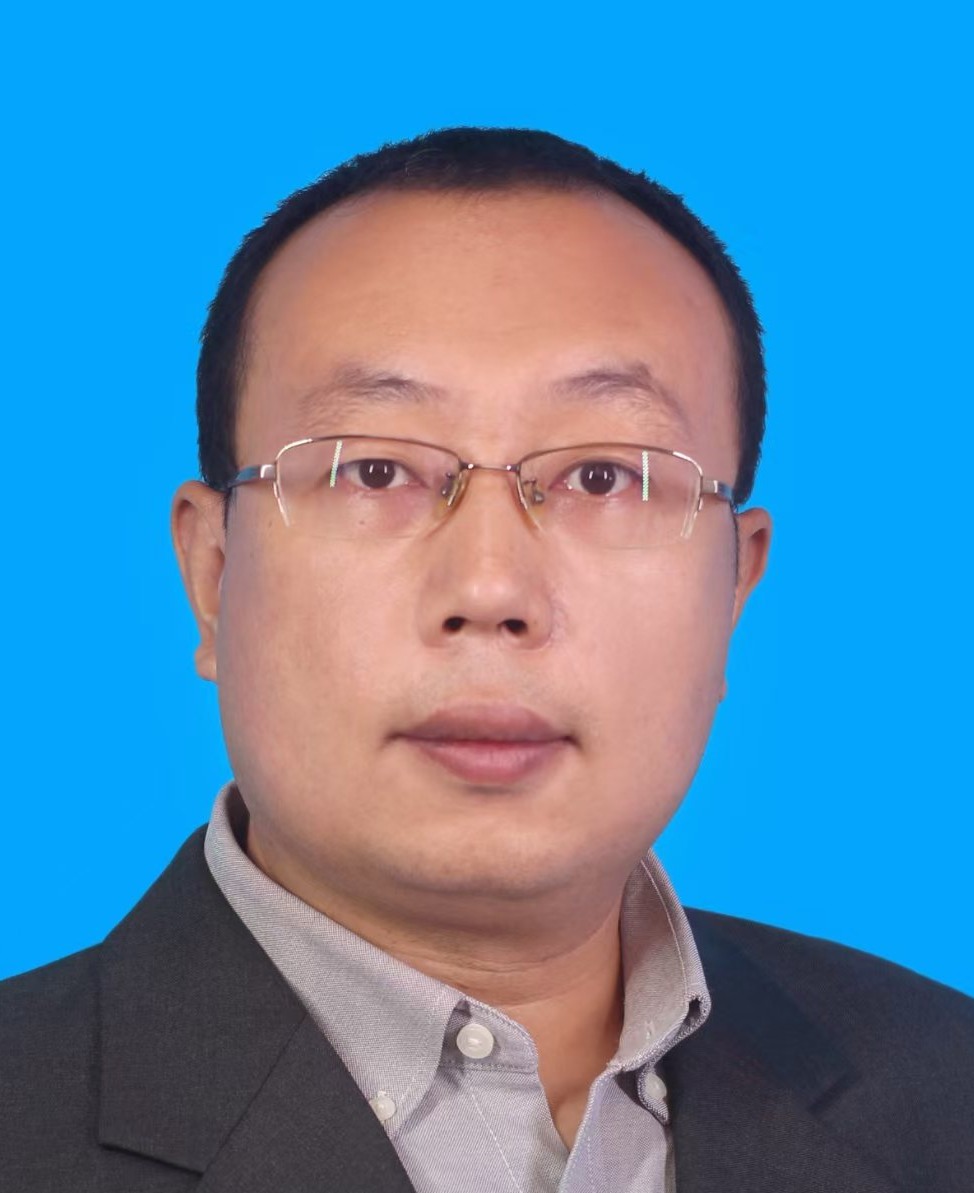}}]{Shuqing Ma} received the Ph.D. degree in underwater acoustic engineering from Harbin Engineering University, Harbin, China, in 2011. He is currently an Associate Professor with the College of Meteorology and Oceanography, National University of Defense Technology, Changsha, China. His research interests include underwater acoustics, underwater acoustic signal processing, and intelligent information processing of underwater multi-physical fields.
\end{IEEEbiography}
\vspace{-33pt}
\begin{IEEEbiography}[{\includegraphics[width=1in,height=1.25in,clip,keepaspectratio]{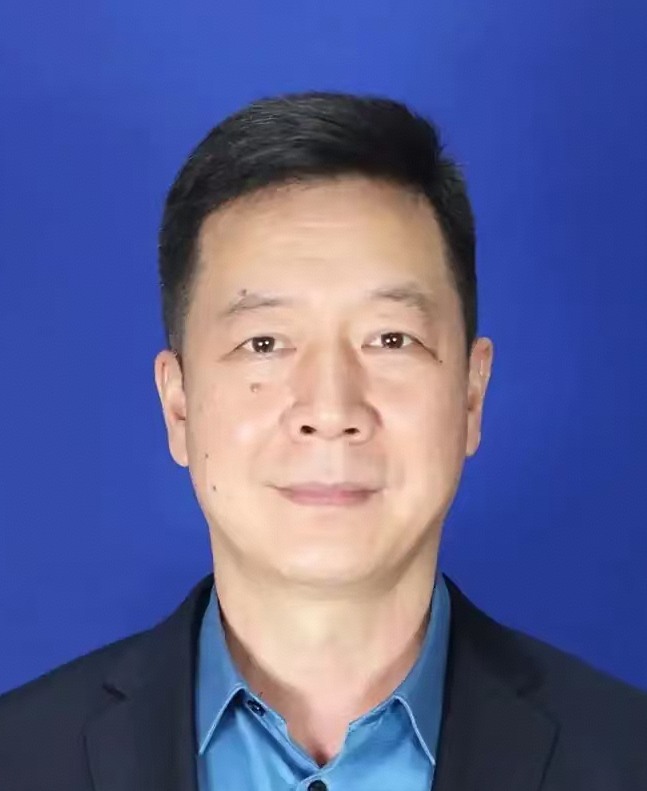}}]{Xiaoqian Zhu} received the Ph.D. degree in computer science and technology from National University of Defense Technology, Changsha, China, in 2007. He is currently a Professor and Doctoral Supervisor with the College of Meteorology and Oceanography, National University of Defense Technology. His research interests include numerical weather prediction, ocean information processing, and underwater target detection. He has led or participated in more than 30 major research projects, including the development of the Global Medium-Range Numerical Weather Prediction System. 
\end{IEEEbiography}

\vfill

\end{document}